\documentclass[authoryear,5p,times,twocolumn]{elsarticle}

\usepackage{hyperref}
\usepackage{graphicx}
\usepackage[caption=false]{subfig}
\usepackage{amsmath}
\usepackage{multirow}
\usepackage{colortbl}
\usepackage{xcolor}
\usepackage{textcomp}

\DeclareMathOperator{\arcsec}{arcsec}
    
\definecolor{mycolor}{rgb}{0.67, 0.88, 0.69}

\journal{Astronomy and Computing}

\biboptions{authoryear}

\begin{document}

\begin{frontmatter}

\title{Machine and Deep Learning Applied to Galaxy Morphology - A Comparative Study}





\author[lac,brandeis]{P. H. Barchi\corref{cor1}}
\cortext[cor1]{Corresponding author}
\ead{paulobarchi@gmail.com}
\ead[url]{paulobarchi.github.io}
\author[nat,astro]{R. R. de Carvalho}
\author[lac]{R. R. Rosa}
\author[lac]{R. A. Sautter}
\author[brandeis]{M. Soares-Santos}
\author[ic]{B.A.D. Marques}
\author[ic]{E. Clua}
\author[valongo]{T. S. Gon\c{c}alves}
\author[valongo]{C. de S\'{a}-Freitas}
\author[iag]{T. C. Moura}

\address[lac]{Lab for Computing and Applied Mathematics, National Institute for Space Research (INPE)\\Av. dos Astronautas, 1.758 - Jardim da Granja, S\~{a}o Jos\'{e} dos Campos - SP 12227-010, Brazil}

\address[brandeis]{Physics Department, Brandeis University}

\address[nat]{NAT - Universidade Cruzeiro do Sul / Universidade Cidade de S\~ao Paulo}

\address[astro]{Astrophysics Division, National Institute for Space Research (INPE)}

\address[ic]{Institute of Computing, Federal Fluminense University (UFF)}

\address[valongo]{Valongo Observatory, 
Federal University of Rio de Janeiro (UFRJ)}

\address[iag]{Instituto de Astronomia, Geof\'isica e Ci\^encias Atmosf\'ericas (IAG), S\~ao Paulo University (USP)}

\begin{abstract}
Morphological classification is a key piece of information to define samples of galaxies aiming to study the large-scale structure of the universe. In essence, the challenge is to build up a robust methodology to perform a reliable morphological estimate from galaxy images. Here, we investigate how to substantially improve the galaxy classification within large datasets by mimicking human classification. We combine accurate visual classifications from the Galaxy Zoo project with machine and deep learning methodologies. We propose two distinct  approaches for galaxy morphology: one based on non-parametric morphology and traditional machine learning algorithms; and another based on Deep Learning. To measure the input features for the traditional machine learning  methodology, we have developed a system called CyMorph, with a novel non-parametric approach to study galaxy morphology.  The main datasets employed comes from the Sloan Digital Sky Survey Data Release 7 (SDSS-DR7). We also discuss the class imbalance problem considering three classes. Performance of each model is mainly measured by Overall Accuracy (OA). A spectroscopic validation with astrophysical parameters is also provided for Decision Tree models to assess the quality of our morphological classification. In all of our samples, both Deep and Traditional Machine Learning approaches have over 94.5\% OA to classify galaxies in two classes (elliptical and spiral). We compare our classification with state-of-the-art morphological classification from literature. Considering only two classes separation, we achieve 99\% of overall accuracy in average when using our deep learning models, and 82\% when using three classes. We provide a catalog with 670,560 galaxies containing our best results, including morphological metrics and classification.
\end{abstract}

\begin{keyword}
galaxies: photometry \sep methods: data analysis \sep machine learning \sep techniques: image processing \sep galaxies: general \sep catalogs
\end{keyword}
\end{frontmatter}

\section{Introduction} 
\label{sec:intro}

In observational cosmology, the morphological classification is the most basic information when creating galaxy catalogs. The first classification system, by \citet{Hubble1,Hubble2}, distinguishes galaxies with dominant bulge component -- also known as Early-Type Galaxies (ETGs) -- from galaxies with a prominent disk component -- named Late-Type Galaxies (LTGs). LTGs are commonly referred to as spiral galaxies because of their prominent spiral arms, while ETGs are commonly referred to as elliptical (E) galaxies as they have a simpler ellipsoidal structure, with less structural differentiation (less information). More refined classifications fork spirals into two groups: barred (SB) and unbarred (S) galaxies. These two groups can also be refined even further by their spiral arms strength. A number known as T-Type can be assigned to the morphological types: ETGs have T-Type $\le$ 0 and LTGs have T-Type $>$ 0 \citep{Vaucouleurs}. T-Type considers ellipticity and spiral arms strength but does not reflect the presence or absence of the bar feature in spirals.

Morphology reveals structural, intrinsic and environmental properties of galaxies. In the local universe, ETGs are mostly situated in the center of galaxy clusters, have a larger mass, less gas, higher velocity dispersion, and older stellar populations than LTGs, which are rich star-forming systems \citep{roberts,blanton,pozzetti}. By mapping where the ETGs are, it is possible to map the large-scale structure of the universe. Therefore, galaxy morphology is of paramount importance for extragalactic research as it relates to stellar properties and key aspects of the evolution and structure of the universe. 

Astronomy has become an extremely data-rich field of knowledge with the advance of new technologies in recent decades. Nowadays it is impossible to rely on human classification given the huge flow of data attained by current research surveys. New telescopes and instruments on board of satellites provide massive datasets.  Therefore, in view of their voluminous size, much of the data are never explored. The potential extraction of knowledge from these collected data is only partially accomplished, even though many answers of the contemporary science critically depend on the processing of such large amount of data \citep{advML4astro,statisticsAstroML,statChallengesInAstro}.
Automatic classification can address this bottleneck of observational research.

One of the most used astronomical datasets is the Sloan Digital Sky Survey -- SDSS, which has been acquiring photometry from the northern sky since 1998. After its first two phases, SDSS Data Release 7 has publicly released photometry for 357 million unique sources, and it is expected to be around 15 terabytes of data when the survey is complete \citep{sdss}. This massive dataset is just one of hundreds of surveys that are currently underway.

One effort to overcome the challenge to classify hundreds of thousands of galaxies depends on the laborious engagement of many people interested in the subject. Galaxy Zoo is a citizen science project which provides a visual morphological classification for nearly one million galaxies in its first phase (Galaxy Zoo 1) distinguishing elliptical from spiral galaxies. With general public help, this project has obtained more than $4 \times 10^7$ individual classifications made by $\sim 10^5$ participants. In its second phase, Galaxy Zoo 2 extends the classification
into more detailed features such as bars, spiral arms, bulges, and many others, providing a catalog with nearly 300 thousand galaxies present in SDSS. Throughout this work, we use Galaxy Zoo \citep{GZ1a,GZ1b,GZ2} classification as supervision and validation (ground truth) to our classification models.

Several authors \citep{abraham,Asymmetry,conselice,lotz} studied and presented results about objective galaxy morphology measures with Concentration, Asymmetry, Smoothness, Gini, and M20 (CASGM system). 
\citet{ferrari} introduced the entropy of information H (Shannon entropy) to quantify the distribution of pixel values in the image. \citet{gpa2018} introduced the Gradient Pattern Analysis (GPA) technique to separate elliptical from spiral galaxies by the second moment of the gradient of the images. This whole system used by \citet{gpa2018} -- called CyMorph -- is described in this paper (Section \ref{sec:cymorph}).  

It is not trivial to determine the success of each non-parametric morphological parameter to perform this classification task. Considering the separation between elliptical and spiral galaxies, for example, 
a morphological parameter is more reliable
if it maximizes the separation of the distributions of these two types.
\citet{gpa2018} described the evaluation technique proposed and adopted to measure the success of metrics to separate elliptical from spiral galaxies \citep[][see also Subsection \ref{sub:ghs}]{pyGHS}. 

\begin{figure*}
  \centering
  \includegraphics[width=0.65\textwidth]{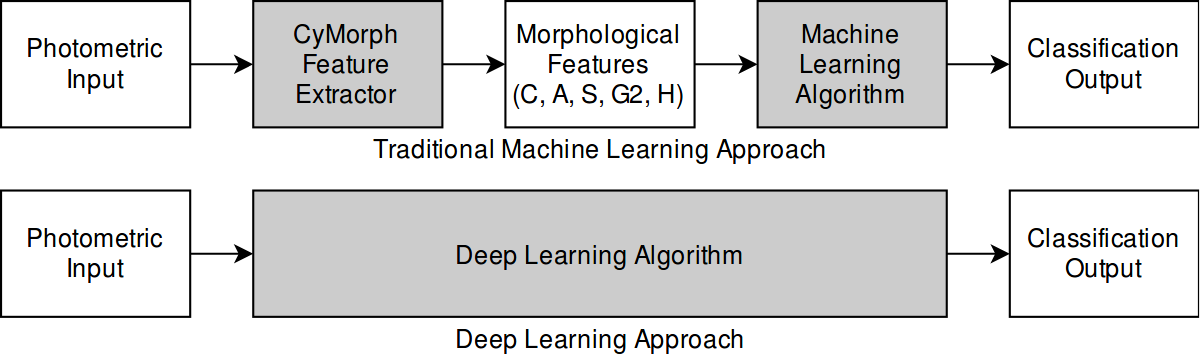}
  \caption{Illustrative sketch of traditional Machine Learning and Deep Learning flows.}
  \label{fig:ml-dl}
\end{figure*}

The main purpose of this investigation is to answer the question ``How to morphologically classify galaxies using Galaxy Zoo \citep{GZ1a,GZ1b,GZ2} classification through non-parametric features and Machine Learning methods?'' We also apply Deep Learning techniques directly to images to overcome the same challenge and compare results from both approaches. Deep Convolutional Neural Network (CNN) is a well-established methodology to classify images \citep{deepLearning}. Without the need of a feature extractor, the network itself adjusts its parameters in the learning process to extract the features. Figure \ref{fig:ml-dl} shows both flows for each approach used in this work: 
Traditional Machine Learning (TML) and Deep Learning (DL). 

The huge amount of photometric astrophysical data available and the highly increasing advancements on hardware and methods to perform automatic classifications has been leveraging related publications \citep{law,freeman,deepGal,blueNugget,jcis,dieleman,khan2018,huertas,deepGal2}. 
Highlight to \citet{deepGal2} who use questions and answers 
from Galaxy Zoo 2 for replicating the answers from the users, and provide morphology classification by T-Type in their final catalog. 

The approach used in this paper is different from the one used in \citet{deepGal2}. Instead of using questions and answers from Galaxy Zoo 2, we use the classifications and images themselves. Also, we revisit issues not touched upon in previous studies dealing with morphological parameters \citep{abraham,Asymmetry,conselice,lotz}; namely, threshold dependence in the use of the segmented image. We study the impact of that on the parameters that ultimately will be used in the TML approach. 

Although it is already a well-established observation that for perception tasks (which galaxy morphology is) Deep Learning is likely to outperform machine learning models trained on hand engineered features \citep{imagenet}, this subject is in its infancy in galaxy morphology and such comparison of these two approaches have never been presented in the same work in the literature. Also, deep learning methods need huge amounts of data to learn from and huge computational resources to make it effective. Deep learning models can be hard to tune and tame, and the prediction time can take much longer than other models because of the complexity \citep{deepLearning}. The traditional machine learning approach is still relevant. 



This document is organized as follows: Section \ref{sec:data} describes the sample and data used to measure morphology and build the classification models. Section \ref{sec:cymorph} describes the advances in non-parametric galaxy morphological system (CyMorph). Sections \ref{sec:ml} and \ref{sec:dl} describe the basics and methodology
of TML and DL employed, respectively. Section \ref{sec:results} presents the results and validation for all experiments conducted. We compare the final product of this work with state-of-the-art catalogs in Section \ref{sec:catalogs}, followed by a summary in Section \ref{sec:summary}. We present catalog details in \ref{sec:fincat}.

\section{Sample and Data}
\label{sec:data}

This work uses data acquired from the SDSS-DR7 \citep{sdss}
and Galaxy Zoo catalogs \citep{GZ1a,GZ1b,GZ2} for measuring morphology and training
the classification models. 
The samples are composed of galaxies in r-band from SDSS-DR7 in the redshift range
$0.03 < z < 0.1$, Petrosian magnitude in r-band brighter than
17.78 (spectroscopic magnitude limit), and $|b| \ge 30^o$, where
$b$ is the galactic latitude. 

For supervised learning purposes, we consider the defined classification from 
Galaxy Zoo 1 \citep[][GZ1 hereafter]{GZ1a,GZ1b} between
E and S galaxies,
and the classification from Galaxy Zoo 2 \citep[][GZ2]{GZ2} with prefixes in one of
11 following classes: Er, Ei, Ec, Sa, Sb, Sc, Sd, SBa, SBc, SBd. 
Other three different scenarios are explored with GZ2 supervision.
Classification considering 9 classes (same as 
11 classes except that we have one class for all elliptical galaxies united), 7 classes (same as previous but disconsidering the faintest galaxy types: Sd and SBd) and three classes: 
E, S and SB.

We study the impact of different datasets on the training process, varying the number and size of objects in the samples.
We define a parameter $K$ as the area of the galaxy's Petrosian ellipse divided by 
the area of the Full Width at Half Maximum (FWHM).
Equation \ref{eq:k} presents how to calculate $K$, where $R_P$ is the Petrosian radius \citep[see][for more details about $R_P$]{petrosian,sdss}. 
By restricting the samples to a minimum $K$, we limit the number and size of objects in the dataset.
The number of galaxies for the three main samples we explore 
($K \ge $ 5,  $K \ge $ 10 and $K \ge $ 20) are presented in Table \ref{tab:sample}.

\begin{equation}
 \label{eq:k}
 K = \left( \frac{R_P}{\textnormal{FWHM}/ 2} \right)^2
\end{equation}

\begin{table}
 \centering
 \caption{Number of galaxies for the main samples in this work from each database (SDSS, GZ1 and GZ2).}
 \vspace*{1mm}
 \begin{tabular}{c c c c}
    \hline
    \multirow{2}{*}{\textbf{Restriction}} & \multicolumn{3}{c}{\textbf{Number of galaxies in}}\\
    \cline{2-4}
    & \textbf{SDSS} & \textbf{GZ1} & \textbf{GZ2} \\ \hline
    $K \ge 5$ & 239,833 & 104,787 & 138,430 \\ 
    $K \ge 10$ & 175,167 & 89,829 & 110,163 \\ 
    $K \ge 20$ & 96,787 & 58,030 & 67,637 \\ \hline
  \end{tabular}
 \label{tab:sample}
\end{table}


With smaller values of $K$ we have more but smaller objects,
while samples restricted by bigger values of $K$ have less but bigger objects.
To properly check the impact of the number and sizes of objects in the samples, we explore the Deep Learning approach for three classes problem in detail with other restrictions: $K \ge $ 7, $K \ge $ 9, $K \ge $ 11, $K \ge $ 14 and $K \ge $ 17.

For Machine and Deep Learning experiments, we split the datasets from GZ1 e GZ2 into training-validation-test subsets in the proportion 80-10-10. In all experiments, each of these subsets are constrained to the same restriction (the model trained and validated with a subset restricted to $K\ge20$ is also tested with the subset restricted to $K\ge20$). 
We should
keep in mind that the data used in this work, SDSS-DR7, does not have a proper spatial resolution (0.396 $\arcsec$ pixel $^{-1}$) and not adequate PSF FWHM ($\sim$1.5 $\arcsec$). For comparison, the Dark Energy Survey \citep[][DES]{DES}, has a pixel size of (0.27 $\arcsec$) and PSF FWHM of $\sim$0.9. This is why we study the quality of our classification as a function of $K$.

\section{Advances in Non-Parametric Galaxy Morphology - CyMorph} 
\label{sec:cymorph}

Methodologies for computing non-parametric morphological metrics have been presented by several authors \citep{morganci,Kent1985ccd,abraham,takamiya,conselice,lotz,ferrari,gpa2018}. 
In this section, we present CyMorph - a non-parametric galaxy morphology system which determines Concentration (C), Asymmetry (A), Smoothness (S), Entropy (H) and Gradient Pattern Analysis (GPA) metrics.




We perform image preprocessing techniques to ensure consistency and improve feature extraction.
CyMorph achieves this goal in three major steps: producing the galaxy stamp, removing secondary objects, and generating the segmented image.
To remove secondary objects inside the stamp we replace their pixels by the median value of the isophotal level that cross that object. 

Concentration is the only metric we calculate using the clean galaxy stamp
since we want the whole accumulated flux profile of the galaxy. For all other metrics, we use the segmented image as input
which we obtain by applying a mask upon the clean image. The mask is computed by a region growing algorithm \citep{pedrini}. We summarize CyMorph metrics as follows:
\begin{itemize}
	\item Concentration is defined as $C = log_{10} (R_1 / R_2)$, where $R_1$ and $R_2$ are the outer and inner radii, respectively, enclosing some fraction of the total flux \citep{conselice,lotz,ferrari}. We use an optimization process for setting up the best configuration parameters for CyMorph (described in Subsection \ref{sub:optm}). The best configuration by this method is: $C = log_{10} \left( R_{75\%} /  R_{35\%} \right)$.
	\item Asymmetry is measured using the correlation between the original and rotated image: $A = 1 - s(I^0, I^\pi)$, where $I^0$ and $I^\pi$ are the original and the $\pi$-rotated images. The function $s()$ is the Spearman's rank correlation coefficient \citep{press}, which has been proved to be a stable and robust correlation coefficient \citep{rubens}.
	\item Smoothness describes the flux dispersion of an image, namely how the gradient varies over the entire image. This can be measured as the correlation between the original image and its smoothed counterpart \citep{abraham,conselice,ferrari}. We apply the Butterworth filter for smoothing the images. This filter provides the advantage of a continuous adaptive control of the smoothing degree applied to the image \citep[see][for more details]{butter,pedrini,rubens}. We use the the Spearman's rank correlation coefficient to compute smoothness, following the same reasoning as for asymmetry. We define smoothness as $S = 1 - s(I^0, I^s)$, where $I^0$ is the flux intensity of the original image, and $I^s$ is the flux intensity of the smoothed image.
	\item Gradient Pattern Analysis (GPA) is a well-established method to estimate the local gradient properties of a set of points, which is generally represented in a two-dimensional (2D) space \citep{gpa1999,gpa2000,gpa2003}. We use the improved version of GPA developed for galaxy morphology (see \citet{gpa2018} and references therein for more details). 
	\item In digital image processing, the entropy of information, H, \citep[Shannon entropy, ][]{bishop} measures the distribution of pixel values in the image. In galaxy morphology, we expect high values of H for clumpy galaxies because of their heterogeneous pixel distribution, and low H for smooth galaxies \citet[see][for more details]{ferrari,bishop}. 
\end{itemize}
For more specific details about how to compute each of these metrics, see \citet{rubens} and references therein.

\subsection{Geometric Histogram Separation ($\delta_{\rm{GHS}}$)}
\label{sub:ghs}

For a given sample of galaxies, CyMorph measures C, A, S, H and GPA and these parameters depend on some quantities. Our main goal is to choose the best quantities possible for reaching a maximum performance in classifying galaxies. Using an independent morphological classification (from GZ1, e.g.), we have elliptical and spiral distributions for each parameter. All we need is a simple and reliable method to objectively assign a value for the separation between elliptical and spiral distributions. 
Here, we measure the geometric distance between the distributions with the GHS (Geometric Histogram Separation) algorithm \citep[see][for more details]{pyGHS,gpa2018}. 

\subsection{Optimizing Morphological Metrics Configuration}
\label{sub:optm}

\begin{table*}
 \centering
 \caption{Parameter ranges explored in the optimization process. Asymmetry is ommited since it depends only on $d_{\sigma}$. Concentration(*) does not depend on $d_{\sigma}$. }
 \vspace*{1mm}
 \begin{tabular}{c c c c c}
    \hline
    \textbf{Sextractor} & \textbf{C}* & \textbf{S} & \textbf{G$_2$} & \textbf{H} \\ \hline
    \multirow{2}{*}{$0.1 \le d_{\sigma} \le 5.0$} & \multirow{2}{*}{\shortstack{$0.55 \le R_1 \le 0.80$\\$0.20 \le R_2 \le 0.45$}} & \multirow{2}{*}{$0.2 \le c \le 0.8$} & \multirow{2}{*}{\shortstack{$ 0.00 \le m_{tol} \le 0.20$\\$0.01 \le p_{tol} \le 0.04$}} & \multirow{2}{*}{$100 \le \beta \le 250$} \\ 
    & & & & \\ \hline
 \end{tabular}
 \label{tab:optim}
\end{table*}

\begin{figure*}
\centering
\subfloat[Concentration]{%
  \includegraphics[width=0.41\linewidth]{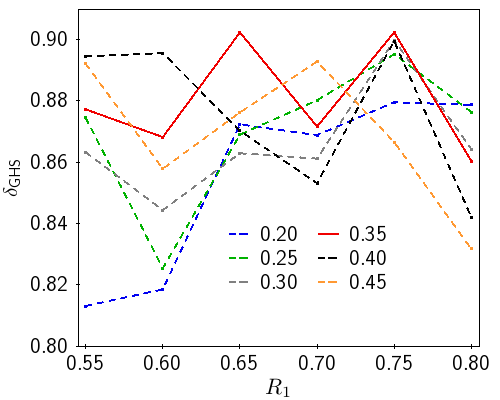}%
  \label{fig:CN}
}%
\subfloat[Asymmetry]{%
  \includegraphics[width=0.41\linewidth]{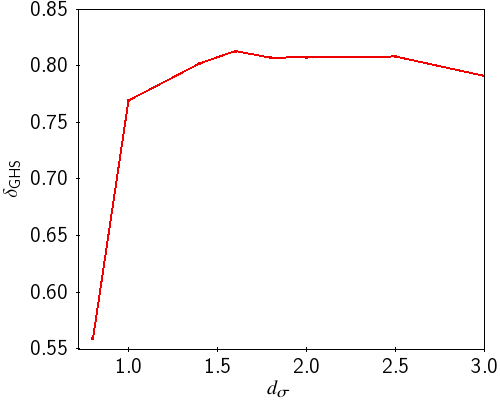}%
  \label{fig:sA3}
}%
\\
\subfloat[Smoothness]{%
  \includegraphics[width=0.41\linewidth]{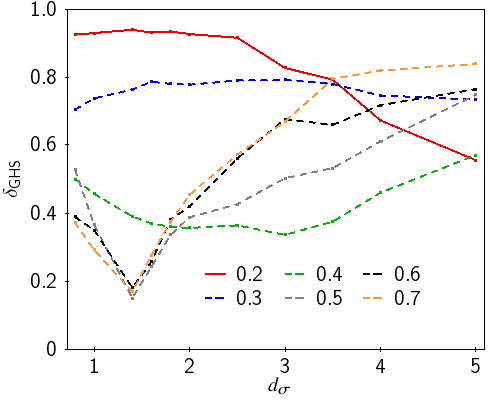}%
  \label{fig:sS3}
}%
\subfloat[Entropy]{%
  \includegraphics[width=0.41\linewidth]{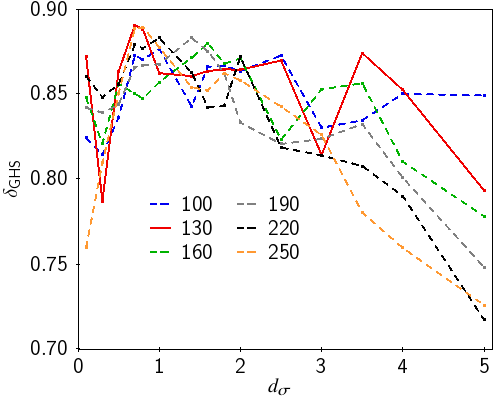}%
  \label{fig:sH}
}%
\\
\subfloat[GPA -- modular tolerance]{%
  \includegraphics[width=0.41\linewidth]{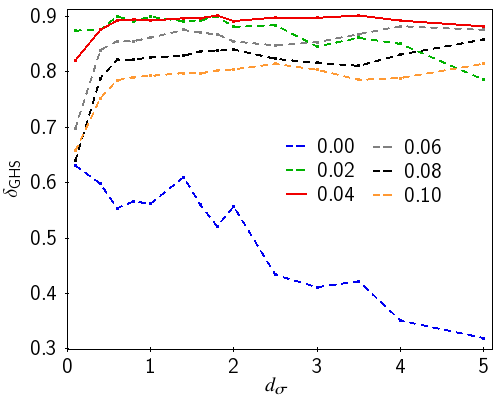}%
  \label{fig:GPA1}%
}
\subfloat[GPA -- fine tuning]{%
  \includegraphics[width=0.41\linewidth]{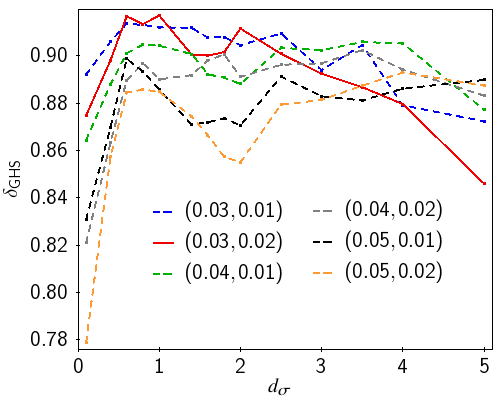}%
  \label{fig:GPA2}%
}
\caption{Plots describing the whole optimization process for morphological metrics 
configuration. Red continuous lines (not dashed) have the best configuration for the 
given parameter. See the explanation for the experiments and best configuration 
results obtained in this Subsection \ref{sub:optm}.}
\label{fig:optim}
\end{figure*}

CyMorph has configurable parameters that we have to fine tune for better distinction between different morphological types. One specific configuration is the threshold parameter used in Sextractor \citep{sextractor} to detect objects on an image: \textit{DETECT\_THRESH} (hereafter $d_{\sigma}$). Sextractor detects an object if a group of connected pixels is above the background by a given $d_{\sigma}$. Thus, we want to find the minimum $d_{\sigma}$ value, sufficiently above the background limit, for which we do not lose information when computing each metric. Most of the configurable parameters are related to each morphological metric. It is important to stress these possibilities to obtain the best performance out of CyMorph. 
Asymmetry only depends on $d_{\sigma}$. For the other metrics, we exhaustively explore the combinations of configurable parameters: outer ($R_1$) and inner ($R_2$) radii for Concentration; control parameter $c$ of Butterworth Filter for Smoothness; modular ($m_{tol}$) and phase tolerance ($p_{tol}$) for G$_{2}$; and, number of bins $\beta$ for Entropy.
Table \ref{tab:optim} summarizes parameters and ranges explored. The optimization process may be approached in different ways. One of them consists of optimizing all variables at once by maximizing a metric which is output from the application of a local two-sample test \citep{kim}. In this work, we focus on a variable-by-variable optimization which not only enables to select the best configuration and input metrics to TML methods but also leaves to the same accuracy in morphology as that obtained in GZ1 (see Section \ref{sec:results}). In the optimization experiments reported here, we randomly select a sample with 1,000 ellipticals and 1,000 spiral galaxies. 
Figure \ref{fig:optim} presents the results for all optimization experiments. In each plot, all lines are dashed except the red one which contains the best configuration for a given metric. The \textit{y}-axis has GHS separation values ($\delta_{\rm{GHS}}$) in every panel. In the following Subsection, we interpret the results displayed in Figure \ref{fig:optim}.

\subsection{Results on Morphology}
\label{sub:CyMorph-results}

\begin{figure}[!h]
  \centering
  \includegraphics[width=.33\textwidth]{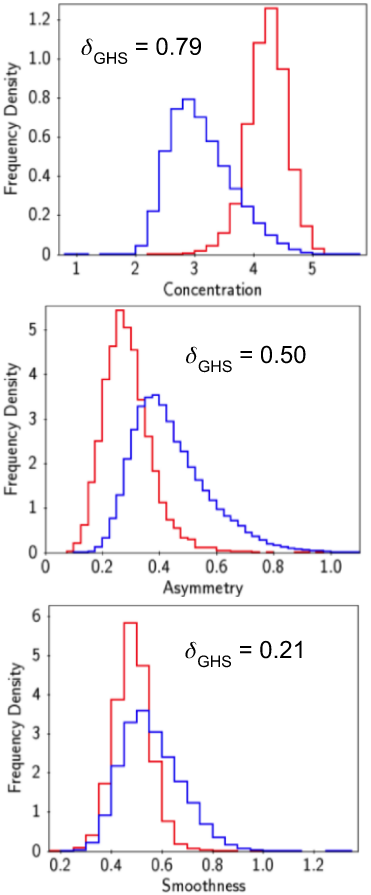}
  \caption{Results on galaxy morphology using Classic CAS \citep{conselice,lotz},
  with elliptical galaxies in red and spiral galaxies in blue.}
  \label{fig:CAS}
\end{figure}

\begin{figure*}
  \centering
  \includegraphics[width=.95\textwidth]{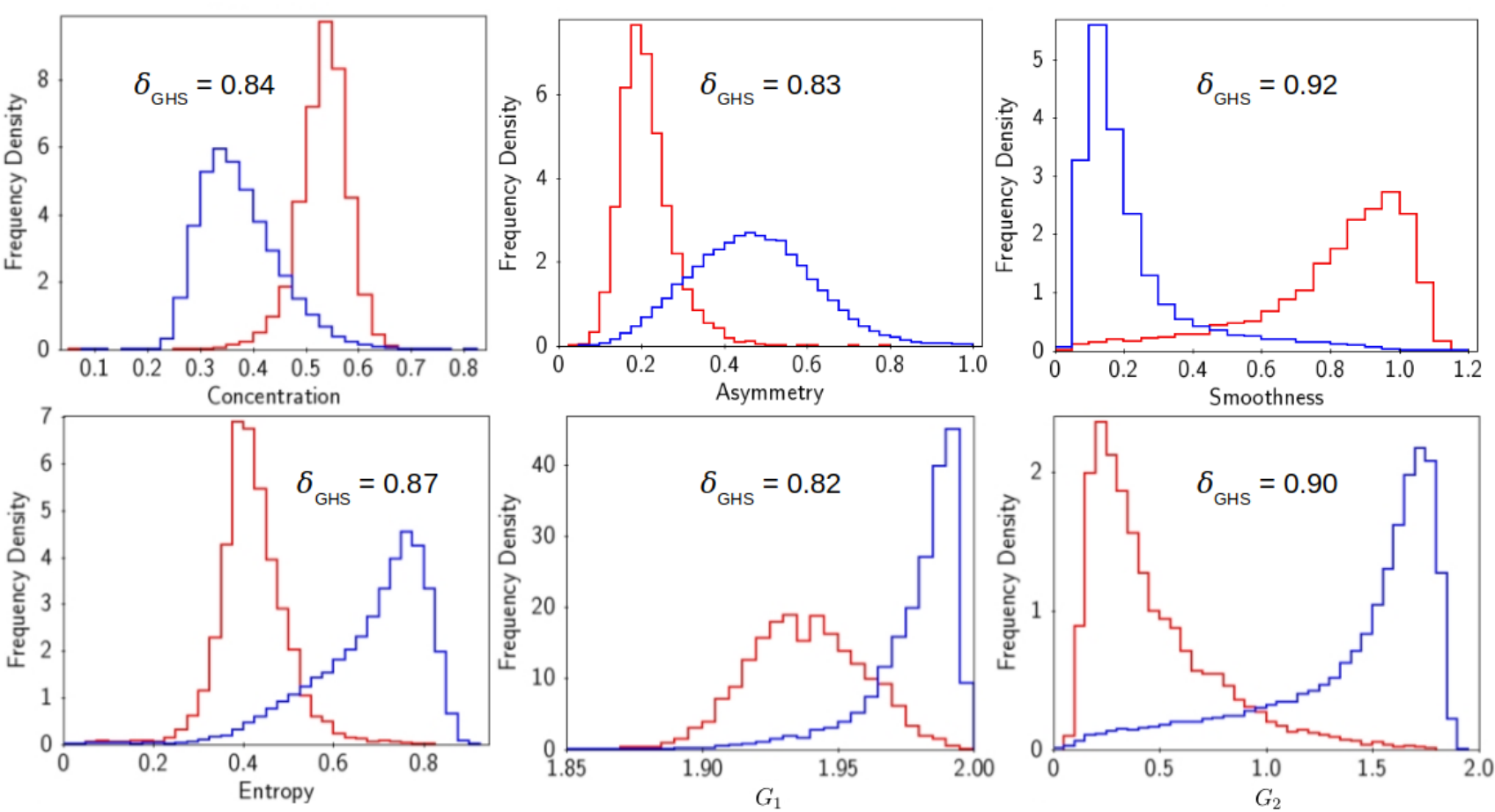}
  \caption{Results on galaxy morphology using CyMorph (the proposed system), 
  with elliptical galaxies in red and spiral galaxies in blue.}
  \label{fig:results}
\end{figure*}

In this subsection, we compare the results obtained by computing the classic CAS system \citep{conselice,lotz}, which are presented in Figure \ref{fig:CAS} and the optimal results obtained by CyMorph system, exhibited in (Figure \ref{fig:results}). \citet{conselice,lotz} estimate Concentration and Asymmetry without significant differences among them. However, Smoothness is implemented in different ways. We present Smoothness as in \citet{lotz}, which gives the most consistent results. For each non-parametric morphological index, we display a binomial distribution histogram with elliptical galaxies (in red) and spiral galaxies (in blue). In each
panel we also list the $\delta_{\rm{GHS}}$ value.

The classic CAS system has the best result with Concentration ($\delta_{\rm{GHS}} = 0.79$), however, this is still lower than the lowest performance obtained by CyMorph metrics, which is Asymmetry with $\delta_{\rm{GHS}} = 0.83$ (see Figure \ref{fig:results}). With this improvement in CAS metrics within CyMorph (Smoothness, for instance, the best result: $\delta_{\rm{GHS}} = 0.92$), and the adoption of Entropy ($\delta_{\rm{GHS}} = 0.87$) and Gradient Pattern Analysis ($G_2$: $\delta_{\rm{GHS}} = 0.90$), we have satisfactory non-parametric morphology metrics to serve as input features to the Traditional Machine Learning algorithms. $G_2$ and $H$, two of the best metrics by $\delta_{\rm{GHS}}$, have highly correlated results: the greater the Entropy value, the more asymmetric gradient patters, and vice versa, lower entropy values correspond to more symmetric gradient patters.

The reasons for the improvement upon classic metrics can be summarized as: (1) the three-step preprocessing, (2) Butterworth filter to smooth the image (concerning Smoothness metric), (3) usage of correlation coefficients for Asymmetry and Smoothness, and (4) optimization process to better configure each metric.

\section{Machine Learning Applied to Galaxy Morphology}
\label{sec:ml}

CyMorph presents a consistent non-parametric morphological system. By employing Machine Learning (ML) methods with CyMorph metrics as features, we can value the best morphological information and obtain reliable and consistent classification results in galaxy morphology. An alternative would be to test logistic regression and other regression methods, which is beyond the scope of this paper. The five input features for the learning process are the best morphological metrics (given by $\delta_{\rm{GHS}}$) computed by CyMorph: C, A, S, G$_2$ and H. We maintain the restriction related to the area of the galaxies to build up different classification models: (1) $K \ge 5$; (2) $K \ge 10$;  and (3) $K \ge 20$, i.e., the area of the galaxy is at least five (model 1), ten (model 2) and twenty (model 3) times larger than the FWHM area for each corresponding object, respectively. 

We build up Decision Tree (DT), Support Vector Machine (SVM) and Multilayer Perceptron (MLP) models to classify galaxies considering different numbers of classes. We use \texttt{scikit-learn} \citep{scikit} \texttt{python} library to perform the experiments and procedures reported in this Section. We use Cross-Validation (CV) to split the dataset in training-validation-testing to address the trade-off between bias and variance \citet{mitchell,hands-on}. First, we split the dataset in a 90/10 proportion for training and testing sets, respectively. CV is applied in the 90\% portion of the dataset. 

Consistent performance validation metrics are crucial to guide the learning process and to objectively measure the performance of each model. No metric is designed to perform this task alone. We employ the Overall Accuracy (OA) as the figure of merit to compare all the different models. 
Additionally, we employ other performance metrics: Precision (P) and Recall (R) -- see \citet{mitchell,hands-on,deepLearning} for more details about OA, P, and R. For a further analysis on the problem with two classes, we use the Receiver Operating Characteristic (ROC) curve and the Area Under the ROC curve \citep[AUC,][]{roc1}.

One of the most used methods for classification and regression is the Decision Tree (DT). 
Among the different versions and variations of DTs, we use the optimized version of Classification and Regression Tree (CART) algorithm. CART builds up binary trees using feature and threshold that yield the largest information gain at each node \citep{tree,hands-on}.


Another influential method for supervised classification is the Support Vector Machine (SVM) which finds the optimal hyperplane that divides the target classes. SVM performs this task by drawing infinite different hyperplanes for separating target classes aiming to get the minimum error rate \citep{svm,svm2}. 



A standard Neural Network (NN) consists of many simple, connected neurons, each one being a computing unit which outputs a sequence of real-valued activations. Neurons are organized in layers: input, hidden (which may be one or many) and output. A Multilayer Perceptron (MLP) has at least three layers (one input, one hidden and one output layer). 
Each neuron has $n$ inputs $i$, weights ($w$), bias ($b$), an activation function ($F(x)$) and output ($y$)
 \citep[see][for more details about NN]{mitchell,hands-on,deepLearning}. We define the MLP architecture by empirically testing different configurations for the two classes problem using the restricted sample defined by $K \ge 20$. We predefine three hidden layers, test different numbers of neurons and alternate between logistic and ReLU as activation functions. Our final NN configuration consists of 44 neurons in the first, 88 in the second and 22 in the third hidden layer, with ReLu as the activation function.

\section{Deep Learning}
\label{sec:dl}

A Neural Network (NN) can be extremely complex when using an architecture with many layers. Deep Learning (DL) methods are built through a deep neural network with multiple layers of non-linear transformations. 
Deep Convolutional Neural Network (CNN) or simply Convolutional Networks \citep{LeCun} are a special kind of neural network for processing data with grid-like topology. 
Convolution preserves the spatial relationship between pixels by using submatrices from the input data to learn features.
It is not feasible to go through all the possibilities of architectures and configurations concerning CNNs. 
In this work, we perform different experiments focusing on two notable robust CNN architectures, Residual Networks \citep[ResNet,][]{resNet} and GoogleNet \citep{GoogleNet}, judging overall accuracy performance and training time. We select the architecture which provide the best overall results: GoogleNet Inception, the winner of ILSVRC 2014 Classification Challenge in visual databases \citep[see][for more details]{GoogleNet,hands-on}.

\section{Results on Classification and Discussion} 
\label{sec:results}

\subsection{Classifier\textquotesingle s performance by Overall Accuracy (OA)}
\label{sub:oa}

\begin{table*}
 \centering
 \caption{Overall Accuracy (OA in percentage) for all approaches considering GZ1 classification
   (elliptical and spiral galaxies separation).}
 \vspace*{1mm}
 \begin{tabular}{c|c|c|c|c||c|c|c|c||c|c|c|c}    
    \multirow{2}{*}{} & \multicolumn{4}{c||}{\textbf{$K \ge 5$}} & \multicolumn{4}{c||}{\textbf{$K \ge 10$}} & 
    \multicolumn{4}{c}{\textbf{$K \ge 20$}} \\ 
    \cline{2-13}
    & \textbf{DT} & \textbf{SVM} & \textbf{MLP} & \textbf{CNN} &
    \textbf{DT}  & \textbf{SVM} & \textbf{MLP} & \textbf{CNN} &
    \textbf{DT}  & \textbf{SVM} & \textbf{MLP} & \textbf{CNN}  \\ \hline
    \textbf{two classes} & 94.8 & 94.6 & 94.6 & 98.7 & 95.7 & 95.8 & 95.6 & 99.1 & 98.5 & 98.6 & 98.6 & 99.5 \\ \hline    
  \end{tabular}
 \label{tab:resultsGZ1}
\end{table*}

\begin{table*}
  \centering
  \caption{Overall Accuracy (OA in percentage) for all approaches considering GZ2 classification. The darker the green colour of a cell, the better OA obtained.}
  \vspace*{1mm}
  \begin{tabular}{c|c|c|c|c||c|c|c|c||c|c|c|c}    
    \multirow{2}{*}{} & \multicolumn{4}{c||}{\textbf{$K \ge 5$}} & \multicolumn{4}{c||}{\textbf{$K \ge 10$}} & 
    \multicolumn{4}{c}{\textbf{$K \ge 20$}} \\ 
    \cline{2-13}
    & \textbf{DT} & \textbf{SVM} & \textbf{MLP} & \textbf{CNN} &
    \textbf{DT}  & \textbf{SVM} & \textbf{MLP} & \textbf{CNN} &
    \textbf{DT}  & \textbf{SVM} & \textbf{MLP} & \textbf{CNN}  \\ \hline
    \textbf{11 classes} & \cellcolor{mycolor!10}49.3 & \cellcolor{mycolor!10}48.8 & \cellcolor{mycolor!10}49.4 & 
    \cellcolor{mycolor!50}63.0 & \cellcolor{mycolor!15}51.6 & \cellcolor{mycolor!15}51.6 & 
    \cellcolor{mycolor!15}51.7 & \cellcolor{mycolor!50}63.0 & \cellcolor{mycolor!25}57.7 & 
    \cellcolor{mycolor!25}57.4 & \cellcolor{mycolor!25}57.7 & \cellcolor{mycolor!50} 65.2 \\ \hline
    \textbf{9 classes}  & \cellcolor{mycolor!50}60.9 & \cellcolor{mycolor!50}63.2 & \cellcolor{mycolor!50}63.0 & 
    \cellcolor{mycolor!95}70.2 & \cellcolor{mycolor!50}60.5 & \cellcolor{mycolor!50}63.8 & 
    \cellcolor{mycolor!50}63.6 & \cellcolor{mycolor!95}75.7 & \cellcolor{mycolor!50}63.5 & 
    \cellcolor{mycolor!50}66.4 & \cellcolor{mycolor!50}66.2 & \cellcolor{mycolor!50}67.4 \\ \hline
    \textbf{7 classes}  & \cellcolor{mycolor!50}63.0 & \cellcolor{mycolor!50}62.5 & \cellcolor{mycolor!50}63.3 & 
    \cellcolor{mycolor!95}72.2 & \cellcolor{mycolor!50}62.9 & \cellcolor{mycolor!50}62.6 & 
    \cellcolor{mycolor!50}63.0 & \cellcolor{mycolor!95}77.6 & \cellcolor{mycolor!50}65.9 & 
    \cellcolor{mycolor!50}65.8 & \cellcolor{mycolor!50}66.0 & \cellcolor{mycolor!95}70.0 \\ \hline
    \textbf{3 classes}  & \cellcolor{mycolor!95}71.9 & \cellcolor{mycolor!95}71.2 & \cellcolor{mycolor!95}71.2 & 
    \cellcolor{mycolor!135}80.8 & \cellcolor{mycolor!95}71.9 & \cellcolor{mycolor!95}74.6 & 
    \cellcolor{mycolor!95}74.9 &\cellcolor{mycolor!145}81.8 & \cellcolor{mycolor!115}78.7 & 
    \cellcolor{mycolor!115}78.5 & \cellcolor{mycolor!115}78.8 & \cellcolor{mycolor!145}82.7 \\ \hline
  \end{tabular}
  \label{tab:resultsGZ2}
\end{table*}

As we have shown in previous sections, there are several parameters driving the final classification
and an appropriate figure of merit is needed to establish which setup/method works best. In Tables \ref{tab:resultsGZ1} and \ref{tab:resultsGZ2}, we present the Overall Accuracy (OA) achieved by all the experiments carried-out in this work. The main goal here is to distinguish between an Early-Type Galaxy (ETG), elliptical (E), and a Late-Type Galaxy (LTG), spiral (S). In the case of TML, using the $K \ge 20$ sample, all methods reached over 98\% of OA. In this training set, there are many more S galaxies ($\sim $87\%) than E galaxies ($\sim$13\%). This difference in the number of examples between classes is called class imbalance. We discuss class imbalance in Subsection \ref{sub:imbalance}. Despite of the imbalance, we have at least 95\% precision and 96\% recall for E systems. Since most of the training set is constituted by S galaxies, it is not surprising that we reach $\sim $99\% precision and recall, establishing a model with $\sim $99\% OA for this dataset.

Overall, CNN is the best approach to establish morphological classification of galaxies. We can safely assert that starting from the classes E and S from Galaxy Zoo 1, we can reproduce the human eye classification with all methods and samples (OA $>$ 94.5\%). When trying to distinguish among 11 classes, the problem is much more complex, as it would be for the human eye, and the best result is $\rm OA \sim 65.2$\% using CNN with $K \ge 20$. However, if we use only three classes we find an OA $>$ 80\% with CNN, for all samples, namely elliptical (E), unbarred (S) and barred spiral (SB) galaxies. We study the three classes problem considering imbalance and different samples in Subsection \ref{sub:imbalance}. 

\subsection{Class Imbalance in Galaxy Morphology}

\label{sub:imbalance}
\begin{figure*}[!h]
\centering
\subfloat[Number of examples as a function of K, for different classes.]{%
  \includegraphics[width=0.32\linewidth]{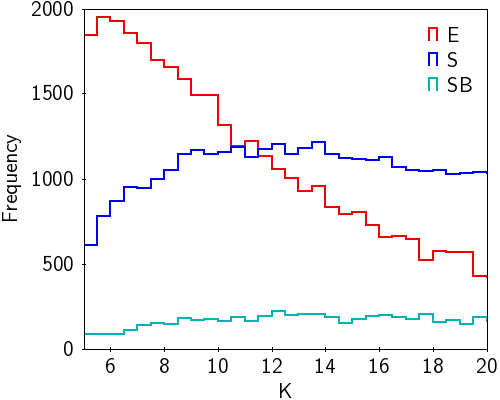}%
  \label{fig:proportions}
}%
\subfloat[Imbalanced (original dataset).]{%
  \includegraphics[width=0.32\linewidth]{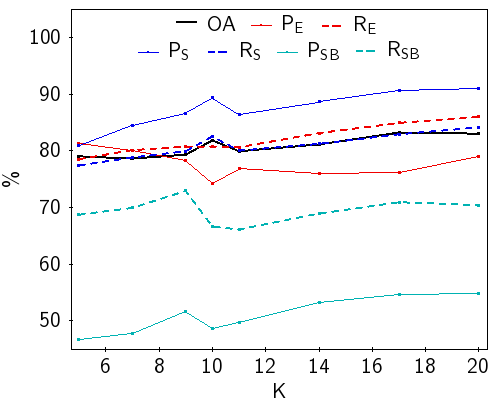}%
  \label{fig:unbalanced}
}%
\\
\subfloat[Balanced -- SMOTE.]{%
  \includegraphics[width=0.32\linewidth]{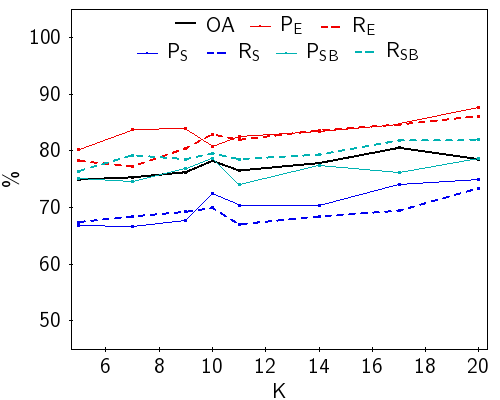}%
  \label{fig:smote}
}
\subfloat[Balanced -- undersampling.]{%
  \includegraphics[width=0.32\linewidth]{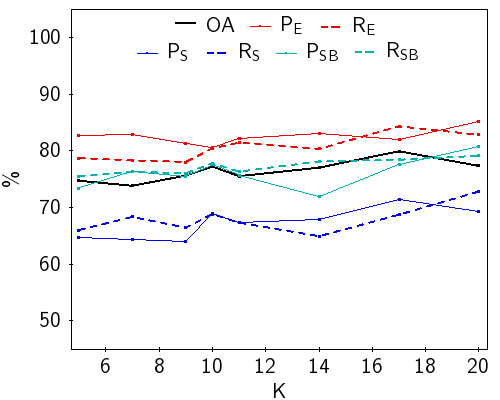}%
  \label{fig:undersampling}
}%
\subfloat[Balanced -- oversampling.]{%
  \includegraphics[width=0.32\linewidth]{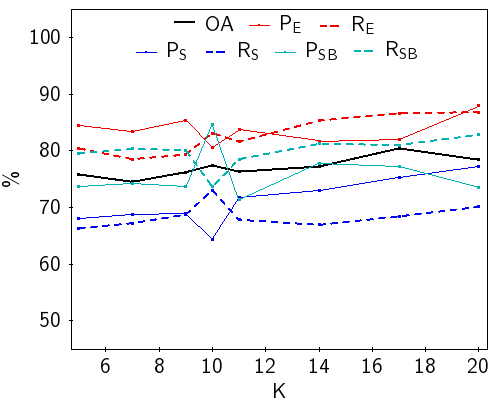}%
  \label{fig:oversampling}
}
\caption{The first plot shows number of elliptical (E), unbarred spiral (S) and barred spiral (SB) galaxies from GZ2 classification varying K. The other four plots are related to the class imbalance problem considering three classes: E (in redder colours), S and SB -- in bluer colours. The black lines indicate the Overall Accuracy (OA). For each of the three classes, continuos lines represent Precision (P) and the dashed lines indicate Recall (R), considering the original imbalanced dataset (panel b), the dataset generated with SMOTE (panel c), the undersampled balanced dataset (panel d) and the oversampled balanced dataset (panel e).}
\label{fig:imbalance}
\end{figure*}

The class imbalance problem is one of the top in data mining, data science, and pattern recognition \citep{10problems}. It arises when at least one of the classes has considerably fewer examples than the other(s). 
This problem is inherent in galaxy morphology, as the number of examples among classes will never be equal. Applying the restriction of $K\ge20$ in the dataset from Galaxy Zoo 1 \citep{GZ1a,GZ1b}, for example, there are  $\sim$ 87\% of galaxies classified as spiral and only $\sim $13\% as elliptical.
Balancing the dataset generally improves the performance for minority classes (since we increase the number of examples of such classes for training), and thus increases precision and recall for these classes \citep{deepLearning}.
Figure \ref{fig:proportions} shows the number of examples from three classes (E, S, SB) in Galaxy Zoo 2 - SDSS DR7 in different bins of K. The bin size is 0.5 and K varies from 5 to 20. SB is the minority class with the number of galaxies approximately constant -- the bar component is a feature identified in all resolutions explored. The number of S galaxies increases until $K = 10$, approximately where the numbers of S and E galaxies are equal. 

We investigate the impact of the imbalance class problem on the morphological classification testing four different datasets: imbalanced, undersampling, oversampling and Synthetic Minority Over-sampling Technique (SMOTE). The imbalanced dataset is the original query. In the undersampling dataset all classes have the same number of examples as SB originally have. For oversampling, we sample the minority class set with replacement. Using SMOTE, we synthetically generate more examples for SB and we consider the smaller of either the number of E galaxies or double the number of SB galaxies to be the number of examples for each class \citep{scikit}.  

Figures \ref{fig:unbalanced}, \ref{fig:smote}, \ref{fig:undersampling} and \ref{fig:oversampling} exhibit OA, P, and R for all experiments exploring the imbalance class problem considering the three classes described above. The minority class (SB) is the one more affected by imbalanced datasets, with low P (51\%) and R (69\%), in average. By employing balancing techniques, we improve to P (76\%) and R (79\%) for the minority class, thus reducing the misclassification for the SB class. All balancing strategies have similar performances. In all strategies, there is a $\Delta \rm{OA} \sim$ 2\% when $K$ varies from 5 to 20. From panels (b) to (e) of Figure 10, we notice that OA weakly increases with $K$, a trend that would imply that restricting the sample to bigger objects reduces classification problems, but the impact is not very significant. Thus, our model built up with the sample restricted by $K \ge$ 5 can safely be used to classify an unknown dataset as it classifies smaller objects with a similar OA compared to bigger objects (such as $K \ge$ 20).

In the remaining of this paper we continue analysing three methods: Traditional Machine Learning (TML) and Deep Learning (DL) approaches using imbalanced dataset for discriminating between two classes; and DL using SMOTE dataset to classify into three classes. For the TML approach, we choose the Decision Tree (DT) algorithm as it is the simplest solution (compared to Support Vector Machine and Artificial Neural Network) and the results have $\Delta \rm{OA} \sim 0$ among them \citep[Occam's razor,][]{occam}. For three classes, we select the model trained with SMOTE dataset because it is in the middle ground between under and overbalancing techniques, and the results using different balanced datasets are equivalent (Figures \ref{fig:smote}, \ref{fig:undersampling} and \ref{fig:oversampling}). These are the selected classifiers to build up the catalog -- see details about the final catalog in \ref{sec:fincat}.

\subsection{Classifier\textquotesingle s performance by ROC curve and AUC}
\label{sub:roc}

\begin{figure*}[!h]
\centering
\subfloat[ROC curves - TML - two classes.]{%
  \includegraphics[width=0.33\linewidth]{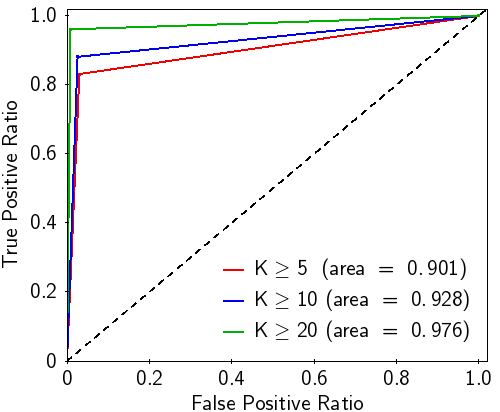}%
  \label{fig:roc-TML-2c}
}%
\subfloat[ROC curves - DL - two classes.]{%
  \includegraphics[width=0.33\linewidth]{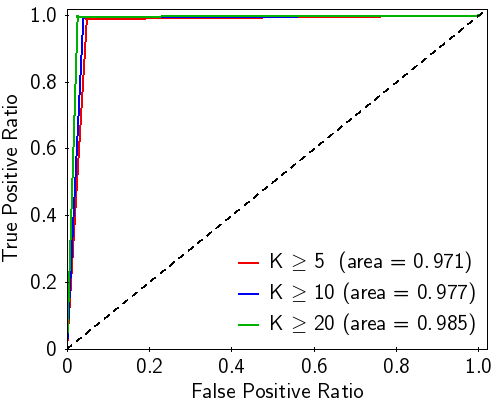}%
  \label{fig:roc-DL-2c}
}%
\\
\subfloat[Truth Probability - TML - two classes.]{%
  \includegraphics[width=0.33\linewidth]{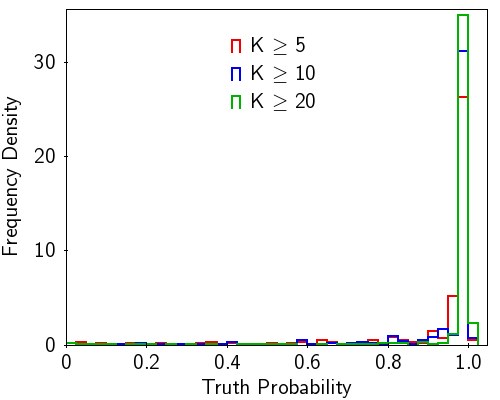}%
  \label{fig:truthProbTML}
}
\subfloat[Truth Probability - DL - two classes.]{%
  \includegraphics[width=0.33\linewidth]{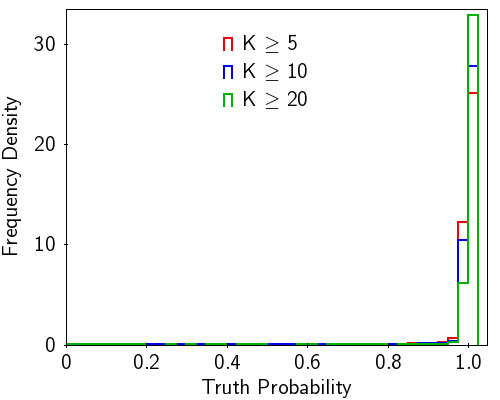}%
  \label{fig:truthProbDL2c}
}%
\subfloat[Truth Probability - DL - three classes.]{%
  \includegraphics[width=0.33\linewidth]{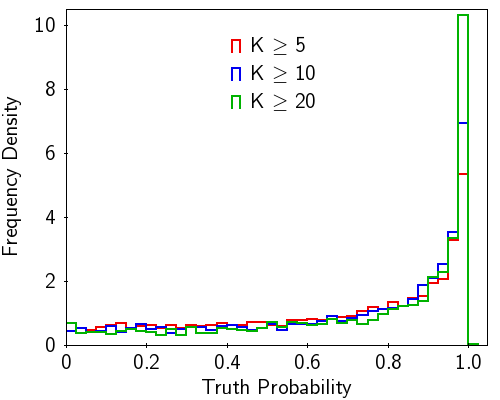}%
  \label{fig:truthProbDL3c}
}
\\
\subfloat[Truth Probability - DL - three classes - $K \ge 5$]{%
  \includegraphics[width=0.33\linewidth]{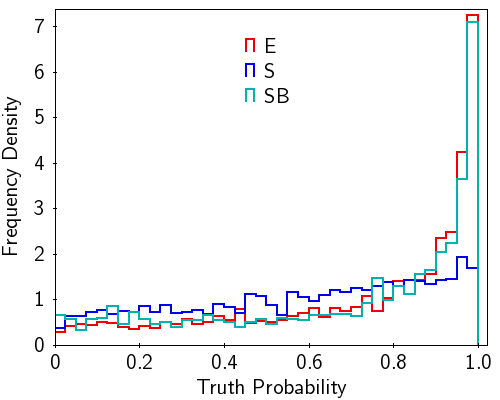}%
  \label{fig:DL-3c-K5}
}
\subfloat[Truth Probability - DL - three classes - $K \ge 10$]{%
  \includegraphics[width=0.33\linewidth]{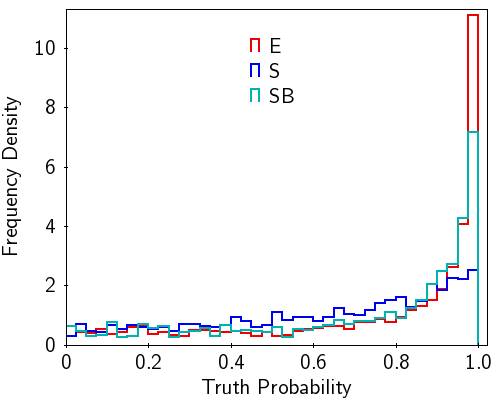}%
  \label{fig:DL-3c-K10}
}%
\subfloat[Truth Probability - DL - three classes - $K \ge 20$]{%
  \includegraphics[width=0.33\linewidth]{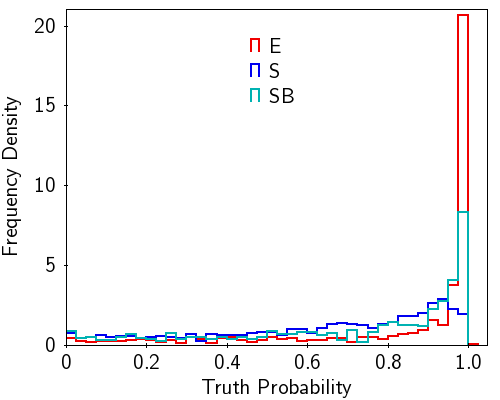}%
  \label{fig:DL-3c-K20}
}
\caption{The first row presents ROC Curve and the Area Under the ROC Curve (AUC -- area) for each approach and different dataset restrictions considering the two classes problem (panels a and b). Such plots consider the ground truth and predicted labels. The dotted black line represents a random guess. The second row shows histograms with ground truth probabilities given by the models for each class (panels c, d, e). The third row presents histograms with ground truth probabilities given by the models for each class in regard to the three classes problem (panels f, g, h).}
\label{fig:roc-curves}
\end{figure*}

One of the most important issues in machine learning is performance measurement. A very popular method is the ROC (Receiver Operating Characteristics) curve and the Area Under the ROC curve \citep[AUC,][]{roc1}. In our particular case, ROC is the probability curve and AUC represents a measure of separability. It indicates how a model is capable of distinguishing between classes. Higher the AUC, better the model is at predicting E's as E's and S's as S's. Based on data presented in Tables \ref{tab:resultsGZ1} and \ref{tab:resultsGZ2}, Figures \ref{fig:roc-TML-2c} and \ref{fig:roc-DL-2c} display the ROC curves. Figures \ref{fig:truthProbTML}, \ref{fig:truthProbDL2c} and \ref{fig:truthProbDL3c} show the histograms with ground truth probabilities given by the models using different datasets, and deeper into the three classes problem Figures \ref{fig:DL-3c-K5}, \ref{fig:DL-3c-K10} and \ref{fig:DL-3c-K20} exhibit histograms with ground truth probabilities given by the models for each class.

ROC curves are typically used in binary classification to study the output of a classifier \citep{roc1,hands-on}. 
Figures \ref{fig:roc-TML-2c} and \ref{fig:roc-DL-2c} show ROC curves considering the ground truth and predicted labels (no probabilities). These ROC curves and area values confirm what Table \ref{tab:resultsGZ1} shows with OA: all models have high standards for acting upon the two classes problem with AUC $>$ 0.90; restricting to the Deep Learning (DL) approach we improve it to AUC $>$ 0.97. By experimenting with different dataset restrictions and approaches we can draw some interesting conclusions. The restriction on the dataset has more impact on TML approach than using DL. The ROC curves are closer to each other on Figure \ref{fig:roc-DL-2c} ($\Delta \rm{AUC} = 0.014$) than \ref{fig:roc-TML-2c} ($\Delta \rm{AUC} = 0.075$). One example is to compare TML using $K \ge 20$ and DL using $K \ge 10$ ($\Delta \rm{OA} \sim 0.5\%$ and $\Delta \rm{AUC} \sim 0$ among them). Using smaller objects, DL can achieve a very similar performance as TML using bigger objects.

The output probabilities given by these models with regard to the ground truth from Galaxy Zoo are explored in Figures \ref{fig:truthProbTML}, \ref{fig:truthProbDL2c}, \ref{fig:truthProbDL3c}, \ref{fig:DL-3c-K5}, \ref{fig:DL-3c-K10} and \ref{fig:DL-3c-K20}). These histograms do not distinguish each class. We consider the output probability from each model for the ground truth of each galaxy.  Again, we confirm that: (1) both approaches have a very high performance considering two classes -- very high concentration of frequency density for truth probability $>$ 0.9, and (2) DL (Figure \ref{fig:truthProbDL2c}) improves the results when comparing to TML (Figure \ref{fig:truthProbTML}) by reducing the frequency density with low truth probability values. The impact of the dataset restriction continues as well: the higher we set the threshold for $K$, the denser the frequency for truth probability $>$ 0.9. 

Although Figure \ref{fig:truthProbDL3c} also presents a high frequency density for truth probability $> 0.9$, it is natural to see a higher frequency density for lower truth probabilities when comparing to Figures \ref{fig:truthProbTML} and \ref{fig:truthProbDL2c} since this problem is more complex when having one more class to consider. Exploring further, Figures \ref{fig:DL-3c-K5}, \ref{fig:DL-3c-K10} and \ref{fig:DL-3c-K20} show different dataset restrictions employed to train models on the three classes problem. Once more, we can see clearly the impact of using bigger (but less) objects: the frequency density for lower truth probabilities decreases from Figure \ref{fig:DL-3c-K5} to Figure \ref{fig:DL-3c-K10} and gets even lower in Figure \ref{fig:DL-3c-K20}. Non-barred spiral galaxies does not have a very high concentration for truth probability $>$ 0.9. However, the other two classes have a high frequency density for truth probability $>$ 0.9. 

\subsection{Learning About Differences between TML and DL From Misclassifications}

\begin{figure*}[!h]
\captionsetup[subfigure]{labelformat=empty,justification=centering}
\centering
\subfloat[587732582054559872\newline GZ1: 0; TML: 1]{%
  \includegraphics[width=0.16\linewidth]{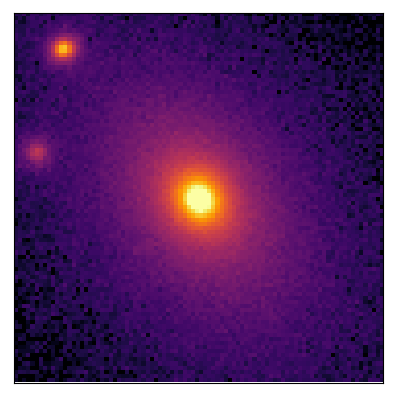}%
}%
\subfloat[587741491440713856\newline GZ1: 0; TML: 1]{%
  \includegraphics[width=0.16\linewidth]{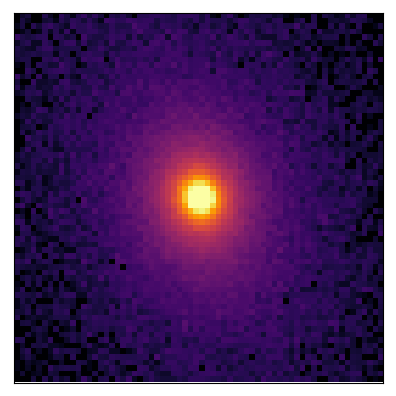}%
}%
\subfloat[588023046942425216\newline GZ1: 0; TML: 1]{%
  \includegraphics[width=0.16\linewidth]{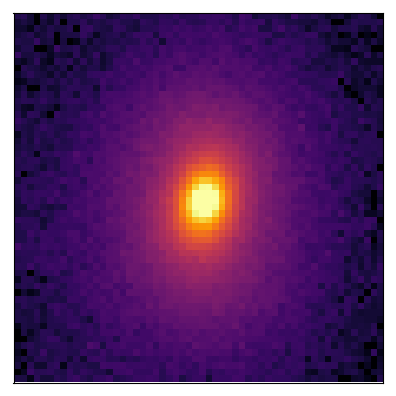}%
}%
\subfloat[587733604255989888\newline GZ1: 1; TML: 0]{%
  \includegraphics[width=0.16\linewidth]{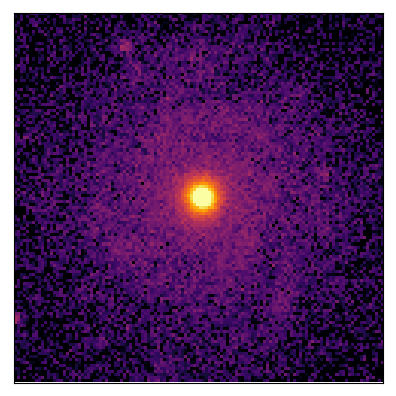}%
}%
\subfloat[587735696448749696\newline GZ1: 1; TML: 0]{%
  \includegraphics[width=0.16\linewidth]{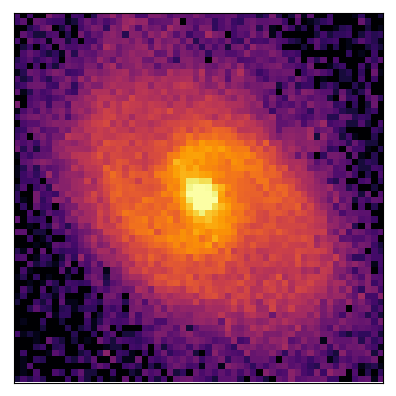}%
}%
\subfloat[587742059994480768\newline GZ1: 1; TML: 0]{%
  \includegraphics[width=0.16\linewidth]{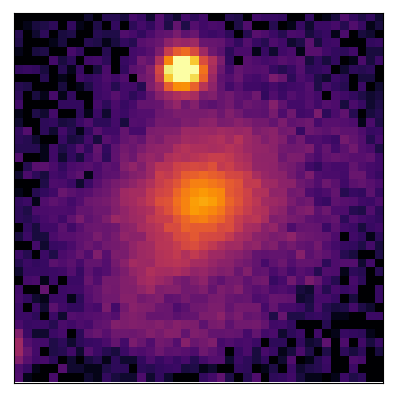}%
}%
\caption{Sample of misclassified galaxies comparing the classification of Galaxy Zoo 1 (GZ1) and our Traditional Machine Learning (TML) approach trained with the sample restricted by $K \ge 20$. Under each galaxy image, we present the object ID number from SDSS-DR7 and the classification given by GZ1 and TML (0: Elliptical; 1: Spiral).}
\label{fig:error_TML}
\end{figure*}

\begin{figure*}[!h]
\captionsetup[subfigure]{labelformat=empty,justification=centering}
\centering
\subfloat[587739305287811105\newline GZ1: 0; DL: 1]{%
  \includegraphics[width=0.16\linewidth]{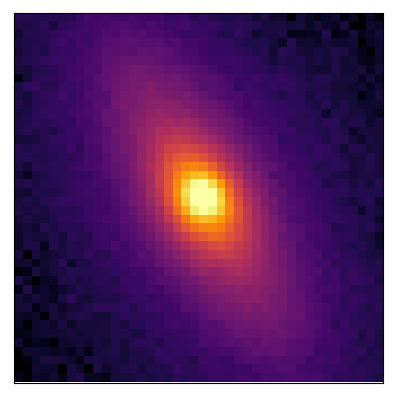}%
}%
\subfloat[587741489822302351\newline GZ1: 0; DL: 1]{%
  \includegraphics[width=0.16\linewidth]{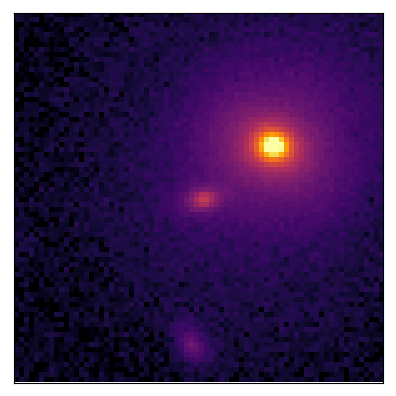}%
}%
\subfloat[588013382210093213\newline GZ1: 0; DL: 1]{%
  \includegraphics[width=0.16\linewidth]{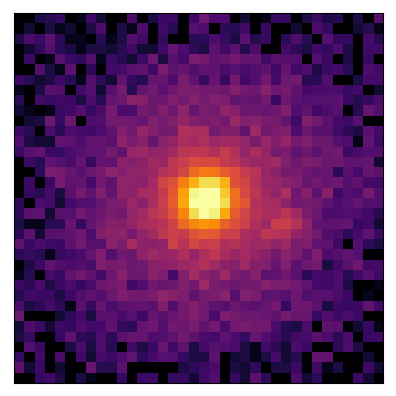}%
}%
\subfloat[587739165702422588\newline GZ1: 1; DL: 0]{%
  \includegraphics[width=0.16\linewidth]{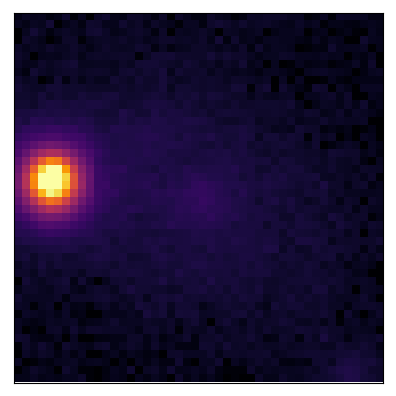}%
}%
\subfloat[587742576444571849\newline GZ1: 1; DL: 0]{%
  \includegraphics[width=0.16\linewidth]{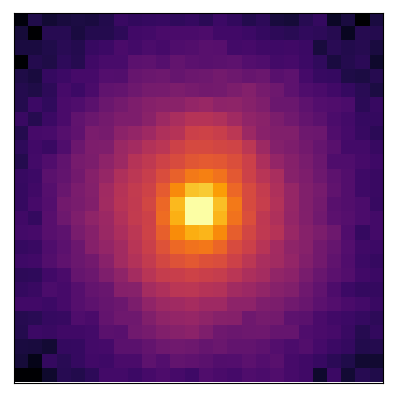}%
}%
\subfloat[588017626688585921\newline GZ1: 1; DL: 0]{%
  \includegraphics[width=0.16\linewidth]{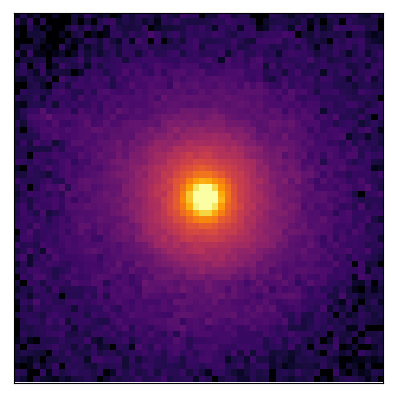}%
}%
\caption{Sample of misclassified galaxies comparing the classification of Galaxy Zoo 1 (GZ1) and our Deep Learning (DL) approach trained with the sample restricted by $K \ge 20$. Under each galaxy image, we present the object ID number from SDSS-DR7 and the classification given by GZ1 and DL (0: Elliptical; 1: Spiral).}
\label{fig:error_DL}
\end{figure*}

The two approaches used in this work, Traditional Machine Learning -- TML; and Deep Learning -- DL, achieve almost the ideal performance considering the Overall Accuracy (OA $\sim$ 99\%) for the two classes problem with the sample restricted by $K \ge 20$. However, it is still worthwhile to further investigate what causes misclassification even at a low percentage. We remind the reader that misclassification is always established using Galaxy Zoo 1 as the ground truth. Figure \ref{fig:error_TML} presents some examples of misclassification using TML. In the first and the last image we see that the preprocessing phase was not able to properly clean the image or discard such examples as bright objects remain close to the central galaxy. The other cases reflect the variance in the parameters used by Decision Tree and the natural uncertainty of the process. In Figure \ref{fig:error_DL} we display some misclassified galaxies by DL. Here, the absence of preprocessing allows galaxies to be too close to the border (second and fourth images) and as before the other examples are simple misclassifications imposed by the method itself, namely the galaxies are easy to be misclassified -- a bright central structure which gradually fades away to the outer part of the galaxy, which, in a more detailed classification (visual), could be considered as a S0 galaxy. We should stress that this misclassification is very low. Using a sample of 6,763 galaxies selected from the grand total of 58,030 galaxies listed in Table 1 ( $K \ge 20$, from GZ1), not used in the training process, TML misclassifies only 72 galaxies (1\%) while DL gets 0.5\% misclassified galaxies. Also, we noticed that none of the galaxies misclassified by TML are in the list of misclassifications by DL. These results seem innocuous, however they remind us of how important is to treat objects close to the border and those near a very bright source as an specific set since this will always be present in any sample. It also reinforces how important it is that we treat independent methodologies along the process of establishing a final morphology attached to an object. The examples presented here show how visual inspection is still an important source of learning about morphology (the problem is not the eye but the quality of the image placed in front of you), although inefficient for large catalogs currently available and the ones coming up in the near future.

\subsection{Validating Classification with Spectroscopic Data}

\begin{figure*}
  \centering
  \includegraphics[width=.95\textwidth]{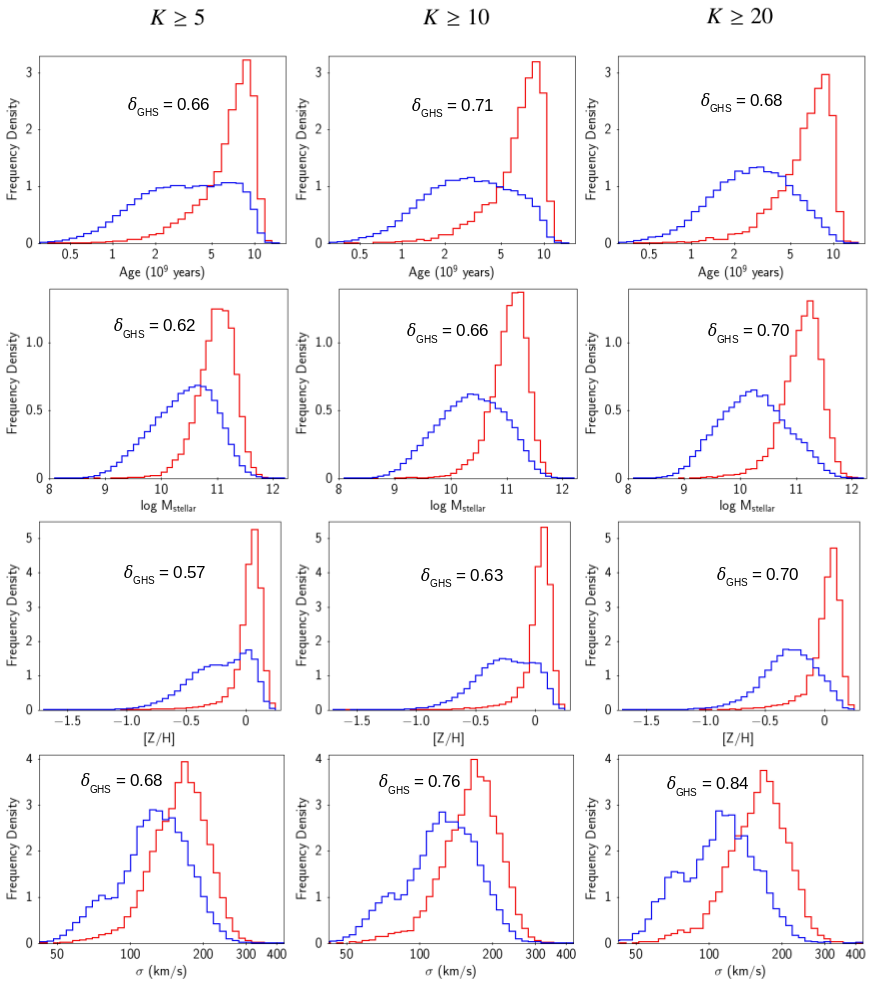}
  \caption{Spectroscopic validation for the ``Undefined'' galaxies from Galaxy Zoo 1 which here are classified by
  our Machine Learning approach using Decision Tree. Elliptical galaxies are displayed in red and spirals in blue. In each panel we give the geometric histogram separation, $\delta_{\rm GHS}$.}
  \label{fig:spectro}
\end{figure*}

The performance analysis presented in the previous section reflects our ability to establish a morphological classification using a given method among several that might in principle works properly, and that's why finding a robust figure of merit is of paramount importance. However, an independent validation is even more essential when presenting a catalog with reliable morphology, namely, we have to show that our new classes recover well know relations. Figure \ref{fig:spectro} presents histograms of Age, stellar mass (M$_{\rm stellar}$), Metallicity ([Z/H]) and central velocity dispersion ($\sigma$) \citep[for more details on how these parameters were obtained and errors, see][]{deCarvalho17}. In every panel we show the distribution for ellipticals (in red) and spirals (in blue). Also, we display the parameter $\delta_{\rm GHS}$ which measures how distant these two distributions are (see Section 3.8). This validation procedure was done using only galaxies from GZ1 classified as "Undefined". The classification here is provided by DT (TML). We remind an important characteristic of the samples: $K \ge 5$, which has more but smaller objects; $K \ge 10$; and $K \ge 20$, which has less but bigger objects. Although we have a bigger dataset with $K \ge 5$, the presence of smaller objects impairs our classification. The degradation of the quality of our classification as we go to smaller galaxies in evident from Figure  \ref{fig:spectro} where $\delta_{\rm GHS}$ decreases for smaller $K$ for all quantities except for Age where only a small fluctuation is noticed.

The number of galaxies for each histogram from Figure \ref{fig:spectro} is as follows:
\begin{itemize}
 \item $K \ge 5$: 13,373 ellipticals; 87,095 spirals; Total: 100,468.
 \item $K \ge 10$: 9,030 ellipticals; 59,096 spirals; Total:   68,126.
 \item $K \ge 20$: 6,390 ellipticals; 24,988 spirals; Total:   31,378.
\end{itemize}

Figure  \ref{fig:spectro} shows how our classification recovers well known properties of galaxies like in the first row we see that ellipticals have ages peaked around 9 Gyr while spirals are younger and the distributions is more spread, probably due to the contamination by S0 galaxies. The second row exhibits the stellar mass distribution and again ellipticals have larger M$_{\rm stellar}$ compared to spirals with a difference of $\sim$0.9 dex, peak to peak. In the third row it is also evident the difference in metallicity ($\sim$0.4 dex), ellipticals are more metal rich than spirals, specially for larger systems. Finally, the distributions of central velocity dispersion show larger values for ellipticals and for spirals we even see a bimodality which reflects the disk to bulge ratio in this morphological type. These distributions attest credibility to our final classification using DT (TML).

\subsection{Case Study: Star Formation Acceleration and Morphologies}
\label{sub:case_study}

In this section we describe an application of the method presented here to study the relation between morphologies and galaxy evolution. More specifically, we use the method to classify a sample of galaxies between disks and spirals and measure quenching timescales for each group separately. There results will be discussed in detail in S\'a-Freitas et al. (in prep).

It has been established that galaxies are show a bimodal distribution in colors, with two distinct peaks with young (blue) and old (red) stellar populations \citep[e.g.,][]{Baldry2004,Wyder2007} and a minimum in the distribution commonly known as the green valley. Although it is generally accepted that galaxies move from blue to red, the physical processes associated with this transition are not completely understood, i.e., we do not know which phenomena are responsible for accelerating the decline in star formation rates, whether a single one or a combination of effects.

Using galaxy colours and stellar population synthesis models, \citet{Schawinski2014} has shown that galaxy quenching can be divided into two distinct processes depending on morphology: elliptical galaxies quench faster, probably through merger activity. \citet{Nogueira-Cavalcante2018} have reached a similar conclusion with more precise measurements from spectral indices (the 4000{\AA} break in the spectral continuum and the equivalent width of the $H_\delta$ absorption line). Nevertheless, both these works rely on assuming specific exponentially declining star-formation histories.

To circumvent this limitation, \citet{Martin2017} have developed a method using the same spectral indices but with no restraints regarding a parametric star formation history. The authors have shown, by comparisons with results from cosmological simulations, that one could infer the instantaneous time derivative of the star-formation rates, denominated the {\it star-formation acceleration}. Formally, this is defined as
\begin{equation}
SFA \equiv \frac{d}{dt}({\rm NUV}-{\rm r}),
\label{eq_sfa}
\end{equation}
with higher values representing stronger quenching.

In S\'a-Freitas et al. (in prep) we apply this methodology to a sample of galaxies out to $z=0.120$, divided by morphology. We only consider galaxies brighter than $M_r=-20.4$ for the sake of sample completeness. When compared to previous works, we are able to measure SFA in galaxies according to morphology for {\it all} objects, regardless of colour and assumed quenching histories. In that sense, the learning techniques presented here are fundamental to our analysis: by classifying a much larger number of galaxies (almost 30,000 galaxies in total), we are able to bin our sample by colours and draw conclusions based on smaller subsamples of objects according to their morphologies.

\begin{figure*}[!h]
\begin{center}
\includegraphics[width=\textwidth]{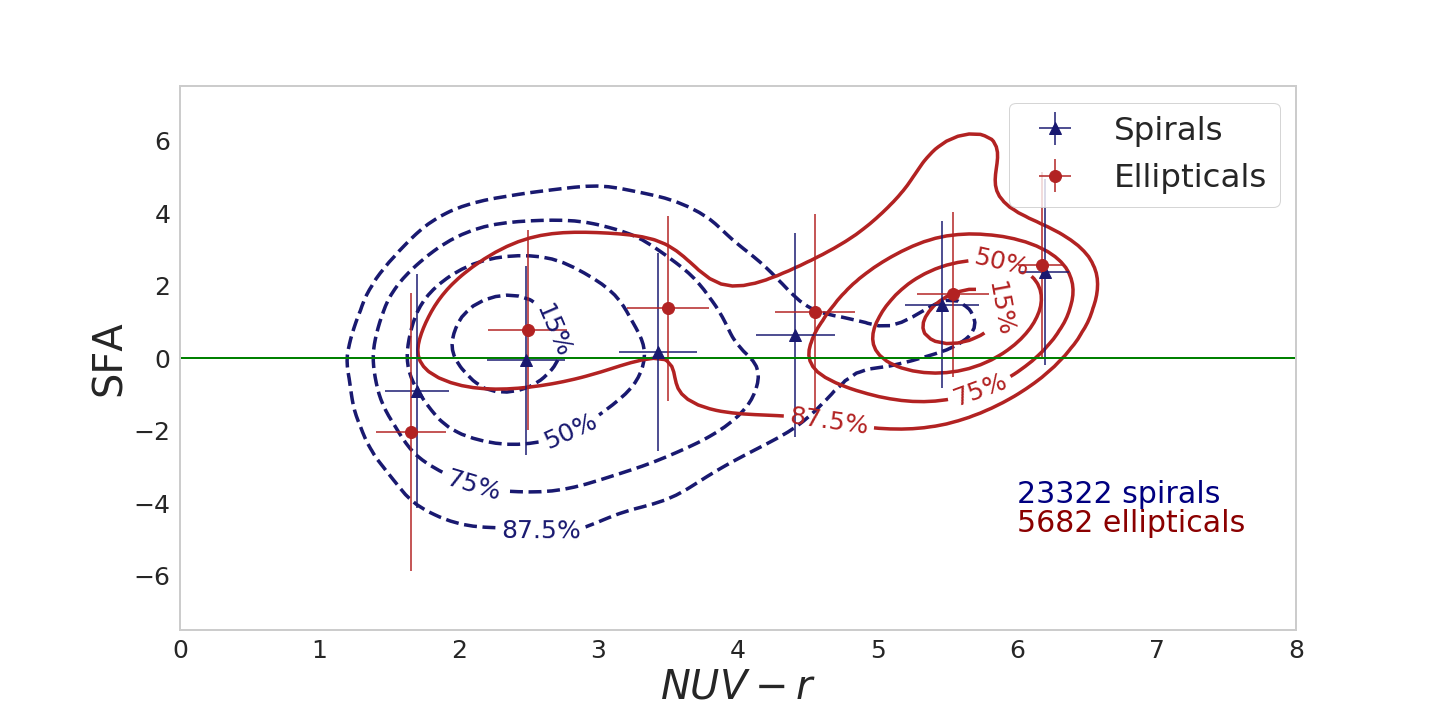}
\caption{Star formation acceleration (SFA) as a function of NUV-r colours. Higher SFA values represent faster quenching, while more negative values indicate strong bursts of star formation, with the green line showing no current variation in SFR. Data are binned in colour, with blue triangles for spiral galaxies and red circles for ellipticals. Error bars show the standard deviation within each bin. Contours show the number of galaxies in the diagram as percentage of the total count for each morphological type. Red galaxies are statistically indistinguishable, while ellipticals in the green valley are quenching significantly faster than spirals. At the blue end, the difference is not large enough for this sample to draw any conclusions.}
\label{camila_sfa}
\end{center}
\end{figure*}

In Figure \ref{camila_sfa} we show our results: as expected, the bluest galaxies are currently undergoing strong bursts, while red galaxies are typically quenching. More importantly, we detect a significant distinction between SFA values for spirals and ellipticals in the green valley. Elliptical galaxies are quenching more strongly, while spirals appear to be moving gradually towards redder colours. We perform Kolmogorov-Smirnoff and Anderson-Darling tests to test for the null hypothesis that the distributions for spirals and ellipticals are drawn from the same parent sample in each bin, ruling this out ($p < 0.05$) only for $2\lesssim ({\rm NUV}-{\rm r})\lesssim 5$. We therefore conclude that this is an effect distinguishable primarily within the green valley, which means that the star formation histories of spirals and ellipticals are only significantly different during their transition to the red sequence.

In the near future, we expect the large upcoming imaging and spectroscopic surveys such as Euclid and DESI to increase our samples significantly, and deep learning techniques will yield reliable morphological classification of millions of objects. This will in turn allow us to further divide our galaxy sample, correlating morphologies with other phenomena such as AGN activity and environment in order to narrow our studies to the specific impact of each on the star formation histories of spiral and elliptical galaxies.

\section{Comparison to Other Available Catalogs}
\label{sec:catalogs}

\begin{figure*}[!h]
\centering
\subfloat[Traditional Machine Learning - two classes.]{%
  \includegraphics[width=0.32\linewidth]{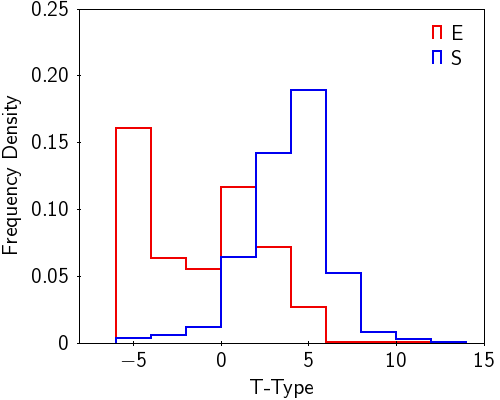}%
  \label{fig:NairTML2classes}
}%
\subfloat[CNN - two classes.]{%
  \includegraphics[width=0.32\linewidth]{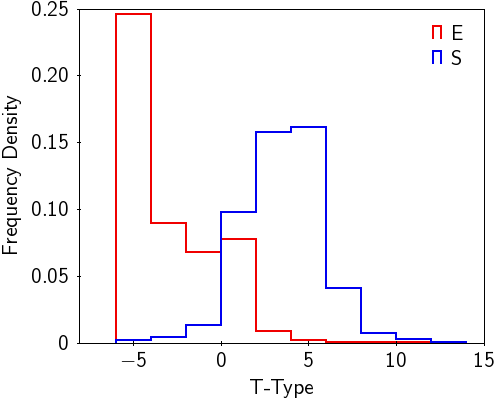}%
  \label{fig:NairCNN2classes}
}%
\subfloat[CNN - three classes.]{%
  \includegraphics[width=0.32\linewidth]{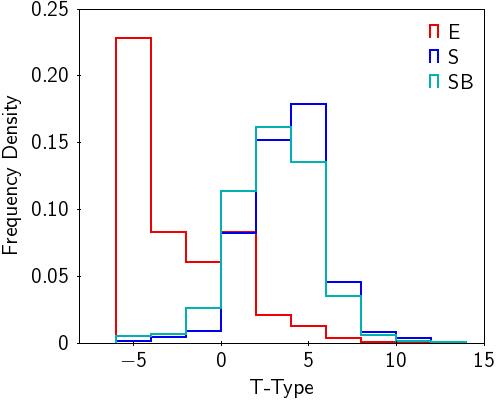}%
  \label{fig:NairCNN3classes}
}%
\caption{Histograms (normalized by area) presenting classifications for \citet{Nair} sample by T-Type
using Traditional Machine Learning classification with two classes
(panel a) and classification with two (panel b) and three classes (panel c) from deep CNN.}
\label{fig:NairClassification}
\end{figure*}

\begin{figure*}[!h]
\centering
\subfloat[Traditional Machine Learning - two classes.]{%
  \includegraphics[width=0.32\linewidth]{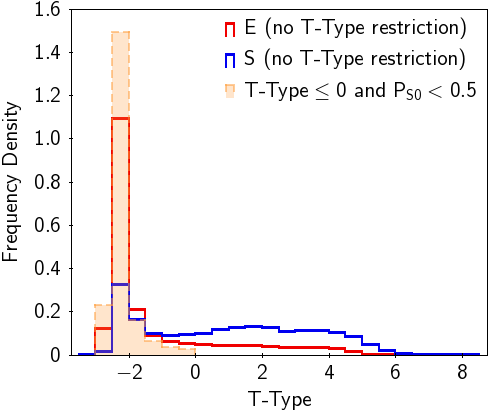}%
  \label{fig:cat-670k-TML2classes}
}%
\subfloat[CNN - two classes.]{%
  \includegraphics[width=0.32\linewidth]{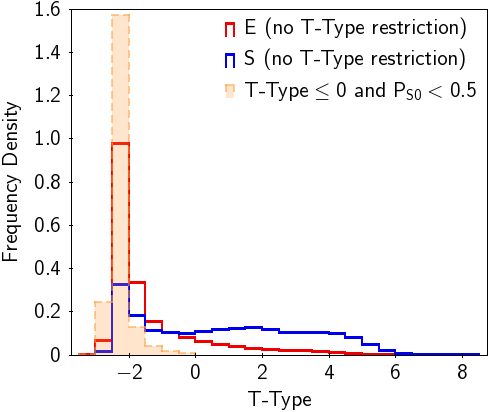}%
  \label{fig:cat-670k-CNN2classes}
}%
\subfloat[CNN - three classes.]{%
  \includegraphics[width=0.32\linewidth]{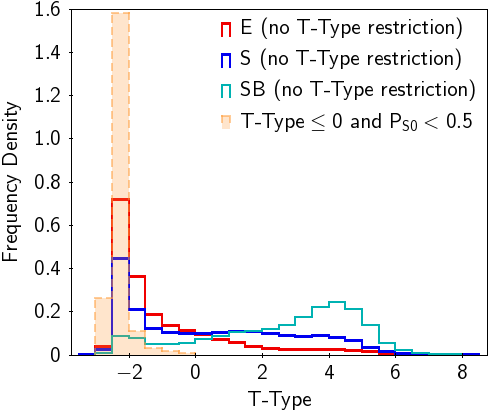}%
  \label{fig:cat-670k-CNN3classes}
}%
\caption{Histograms presenting classifications for \citet{deepGal2} sample by T-Type
using Traditional Machine Learning classification with two classes (panel a), and,
classification with two (panel b) and three classes (panel c) from deep CNN (normalized by area).}
\label{fig:cat-670k-classification}
\end{figure*}

\begin{figure*}[!h]
\captionsetup[subfigure]{labelformat=empty}
\centering
\subfloat[587742903942840576]{%
  \includegraphics[width=0.16\linewidth]{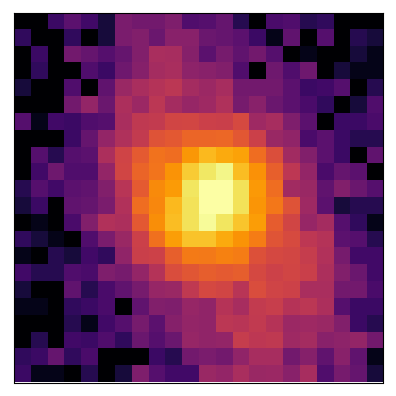}%
}%
\subfloat[587745244699623552]{%
  \includegraphics[width=0.16\linewidth]{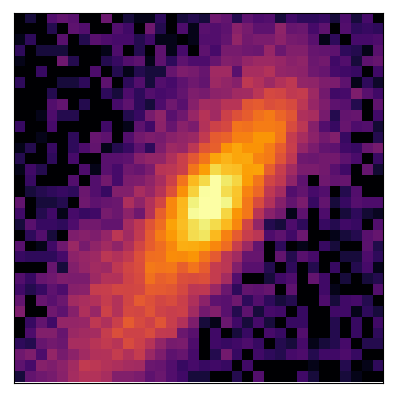}%
}%
\subfloat[588007004732719488]{%
  \includegraphics[width=0.16\linewidth]{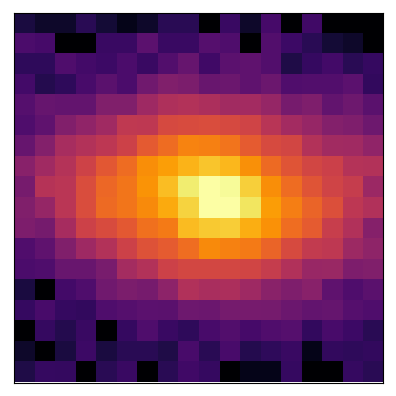}%
}%
\subfloat[588007005231776000]{%
  \includegraphics[width=0.16\linewidth]{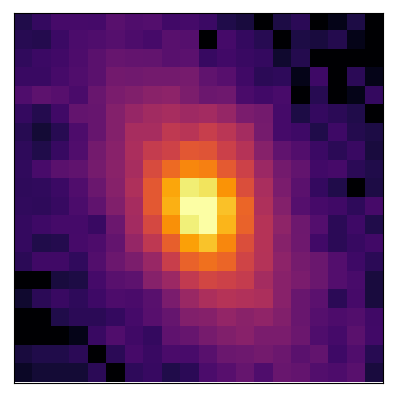}%
}%
\subfloat[588007005239050496]{%
  \includegraphics[width=0.16\linewidth]{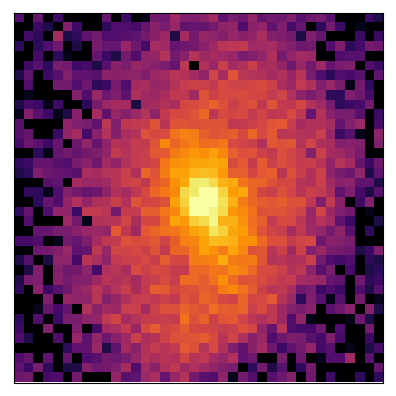}%
}%
\subfloat[588009367478403200]{%
  \includegraphics[width=0.16\linewidth]{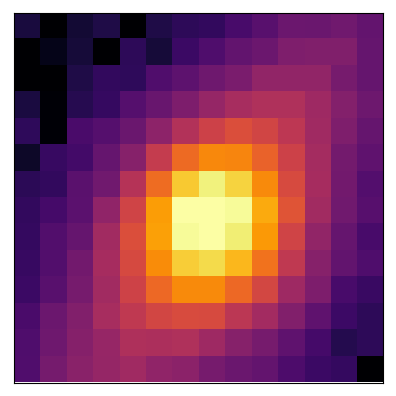}%
}%
\caption{Sample classified as spiral galaxies by our classifier with -2.25 $\le$ T-Type $\le$ -2 by \citet{deepGal2}. The object ID number from SDSS-DR7 is presented under each galaxy image.}
\label{fig:S-TType-sample}
\end{figure*}

To attest the reliability of the morphological classification we provide in this work (see \ref{sec:fincat} for details about our catalog), it is of paramount importance to do external comparisons. There are currently two reliable catalogs that serve this purpose. First, \citet{Nair} provide T-Type information for 14,034 galaxies visually classified by an expert astronomer. Second, \citet{deepGal2} lists
670,722 galaxies also with T-Type available. 

Figure \ref{fig:NairClassification} presents the histogram of T-Type provided by \citet{Nair} for the elliptical and spirals classes from our work. In general, the distributions are as expected - ellipticals peak around T-Type = -5 and spirals around T-Type = 5. In all three cases we notice an extension of the histogram for ellipticals towards larger T-Types, like a secondary peak around T-Type = 1, which may be associated to S0 galaxies.  In Figure \ref{fig:NairCNN2classes}, we note an improvement in using DL over TML by observing the decrease of the fraction of elliptical galaxies and the corresponding increase of spiral galaxies with T-Type $>$ 0. Such behavior is also there in Figure \ref{fig:NairCNN3classes}, considering three classes, with elliptical galaxies mostly with T-Type $\le$ 0 and spirals (S and SB) primarily with T-Type $>$ 0. The general comparison to the classification obtained by \citet{Nair} exhibit a 87\% agreement.

Figure \ref{fig:cat-670k-classification}, analogous to Figure \ref{fig:NairClassification}, shows how our classification
performs in comparison with that provided by \citet{deepGal2}. In all panels, we see a striking different {\it wrt} the comparison with  \citet{Nair} - a considerable amount of spirals around T-Type $\sim$ -2. Along with T-Type, \citet{deepGal2} provide also the probability of each galaxy being S0: $P_{\rm{S0}}$. They define elliptical galaxies as those with T-Type $\le 0$ and $P_{\rm{S0}} < 0.5$; S0 galaxies have T-Type $\le 0$ and $P_{\rm{S0}} > 0.5$; and spiral galaxies have T-Type $> 0$. We plot the elliptical galaxies, following their definition, as a filled histogram in orange, and this shows a higher peak at T-Type $\sim$ -2 {\it wrt} the distribution of the ellipticals with no T-Type restriction. Therefore, restricting the definition we get much higher concordance, namely higher fraction of systems that we classify as ellipticals, which translates into a higher peak around T-Type $\sim$ -2. Not only this, but as we can see from panel (c), using the three classes morphologies, the fraction of ellipticals with no T-Type restriction gets lower and the barred spirals appear more prominently around T-Type $\sim$ 4. In the same way we did when comparing to \citet{Nair}, here we find a 77\% agreement when comparing only elliptical and spiral galaxies. 

A final note on the comparison with \citet{deepGal2} is related to the S0 class, in which we see a prominent bulge and a disk. They classify 230,217 as S0 and 27.96\% of these systems (64,380) have $K < 5$, i.e., $\sim$ 28\% of galaxies classified as S0 are very small objects. Visually, it is easy to misclassify galaxies with predominant, oval and bright structure; and the task becomes even more difficult if the objects are small. Figure \ref{fig:S-TType-sample} shows a sample of galaxies with -2.25 $\le$ T-Type $\le$ -2 according to \citet{deepGal2} that we classify as spiral galaxies. As our classifiers do not discriminate the S0 morphology it is not surprising that we classify as spiral if the galaxy has a prominent disk. 

Finally, we note that since \citet{Nair} and \citet{deepGal2} present their classification as T-Type, a proper comparison is difficult to make. However, the agreement displayed in Figures \ref{fig:NairClassification} and \ref{fig:cat-670k-classification}, together with the global concordance when 
comparing elliptical and spiral galaxies, give us confidence that the classifications obtained here in this work are consistent and robust.

\section{Summary} 
\label{sec:summary}

With the new photometric surveys coming up, in several bands and with varying depth, it is of paramount importance to have the proper machinery for morphological classification, which is one of the first elements to create 
a reliable galaxy catalog, from which we can select clusters of galaxies and study the large scale structure of
the universe. Here, we present models and methodologies to achieve these goals. We investigate the limits of applicability of TML and DL, in the supervised mode and compare their performances. We revisit the non-parametric
methodology using C, A, S, H and G$_{2}$ and study some details ignored in previous works. Also, we examine
how different methods are sensitive to the size of the galaxies, here identified by the ratio between the object's area and the PSF area.  Finally, we remind the readers again of the importance of comparing TML with DL, since they are radically different approaches that in principle should result in similar classes. In the following, we summarize the main contributions of this paper:

\begin{itemize}

\item We investigated how parameters involved in the TML (S,A,H, and G$_{2}$) depend on the threshold used to obtain the segmented image. Although this seems a minor detail, it has proven to be an important ingredient
in improving the TML performance, since the separation between ellipticals and spirals, $\delta_{\rm GHS}$,
is maximized according to the threshold. Comparison with the traditional CAS system shows a considerable
improvement in distinguishing ellipticals from spirals, namely, for CAS ($\delta_{\rm GHS}$ = 0.79, 0.50, 0.21 for
C, A and S respectively), while in our modified CAS we have $\delta_{\rm GHS}$ = 0.84, 0.83, 0.92. Besides,
the new parameters H and G$_{2}$ have their  $\delta_{\rm GHS}$ values very high (0.87 and 0.90, respectively)
attesting their usefulness in galaxy morphology analysis. We list all these parameters in our main catalog with 670,560 galaxies.

\item One way of testing the quality of our morphological classification (based on photometric data) is to compare
with independent classes established with different data (spectroscopy). We use only galaxies from Galaxy Zoo 1 classified as ÓUndefinedÓ and applied our Decision Tree with the traditional machine learning approach. The result
is presented in Figure  \ref{fig:spectro} where we see an excellent performance of our classes in distinguishing
the stellar population properties of ellipticals and spirals. Ellipticals have ages peaked around 9 Gyr while spirals are younger and the distribution is more spread. Ellipticals have larger M$_{stellar}$ compared to spirals with a difference of $\sim$0.9 dex.  The difference in metallicity ($\sim$0.4 dex) between ellipticals and spirals is noticeable, specially for larger systems. Also, ellipticals show larger values of central velocity dispersion show larger values for ellipticals compared to spirals, for which we even see a bimodality reflecting the disk to bulge ratio variation in this morphological type.

\item We present a preliminary result on SFA which study was always hampered by the lack of reliable morphological classification for a sizeable sample. Our catalog provides the necessary input data for such analysis. We show that the bluest galaxies are currently experiencing strong bursts while red galaxies are quenching. Also, we present for the first time a significant distinction between SFA values for spirals and ellipticals in the green valley. We find that the star formation histories of spirals and ellipticals are only significantly different during their transition to the red sequence. A full analysis of this topic and consequences for galaxy evolution is presented in S\'a-Freitas et al. (in prep).

\item We use a deep convolutional neural network (CNN) - GoogLeNet Inception, to obtain morphological classifications for galaxies for all galaxies in the main catalog under study here. With the twenty-two layer network and imbalanced datasets, the results obtained considering two classes are very consistent ($\rm OA\ge98.7\%$) and for the three classes problem they are still good, considering the quality of the data ($\rm OA \sim 82\%$). Also, in comparison with TML, DL outperforms by $\Delta \rm OA \sim 4\%$ and $\Delta \rm AUC \sim 0.07$ for galaxies with  K$\ge$ 5.

\item We make public a complete catalog for 670,560 galaxies, in the redshift range 0.03 $<$ z $<$ 0.1, Petrosian magnitude in r-band brighter than 17.78, and $|b| \ge 30^o$. The input data comes from SDSS-DR7. We provide morphological classification using TML and DL, together with all parameters measured with our new non-parametric method (see \ref{sec:fincat} for catalog details). We append classifications (T-Type) from \citet{Nair} and \citet{deepGal2} whenever available. 
 
\end{itemize}

\section*{Acknowledgements}

This work was financed in part by the Coordena\c{c}\~ao de Aperfei\c{c}oamento de Pessoal de N\'{i}vel Superior - Brasil (CAPES) - Finance Code 001. R.R.dC. and R.R.R. acknowledge financial support from FAPESP through grant \# 2014/11156-4. RRR and PHB thank Santos Dumont Supercomputer-LNCC for  providing 500KUAs for the partial development of this research. The authors thank to MCTIC/FINEP (CT-INFRA grant 0112052700), the Embrace Space Weather Program for the computing facilities at INPE and NVIDIA for providing GPUs. We thank Dr. Diego Stalder, Dr. Bjoern Penning, Alyssa Garcia, Luke Korley, Dr. Helena Dom\'{i}nguez Sanch\'{e}z, and Dr. Sandro B. Rembold for productive discussions and thoughtful comments on several topics related to the present work. 

\appendix
\label{appendix}

\section{Final Catalog}
\label{sec:fincat}

The final product of this work is a catalog with morphological information for 670,560 galaxies. The input data comes from SDSS-DR7 and the sample is restricted by the redshift range 0.03 $<$ z $<$ 0.1, Petrosian magnitude in r-band brighter than 17.78, and $|b| \ge 30^o$. We provide morphological classification using TML and DL approaches for distinguishing elliptical (0) from spiral (1) galaxies. Furthermore, using DL approach, we release classification considering three classes: ellipticals (0), unbarred spirals (1) and barred spirals (2). For DL classification, we exhibit the classes ordered by probability and respective confidence percentages. We provide our best morphological non-parametric parameters as well: Concentration (C), Asymmetry (A), Smoothness (S), Gradient Pattern Analysis parameter (G$_2$) and Entropy (H). The columns we provide are: the value of the parameter K, CyMorph metrics (5 columns), CyMorph Error, TML classification considering two classes, DL classes considering two classes and their respective percentages (4 columns),  DL classes considering 3 classes, and their respective percentages (6 columns). In detail:
\begin{itemize}
 \item $K$ is the area of the galaxy's Petrosian ellipse divided by the area of the Full Width at Half Maximum (FWHM).
 \item $C$, $A$, $S$, $G_2$ and $H$ are the non-parametric morphological parameters from the CyMorph system (see Section \ref{sec:cymorph});
 \item $Error$ contains the Error flag after processing CyMorph;
 \item $ML2classes$ is the classification obtained with the TML approach, using CyMorph and Decision Tree to separate galaxies into elliptical and spiral galaxies. Here, we maintain the restriction about $K$: galaxies with $5 \le K < 10$ are classified by the model built up with the sample with $K \ge 5$ restriction; galaxies with $10 \le K < 20$ are classified by the model built up with the sample with $K \ge 10$ restriction; galaxies with $K \ge 20$ are classified by the model built up with the sample with $K \ge 20$ restriction; galaxies with $K < 5$ are not classified.
 \item $CNN2classes1stClass$ is the class with the highest probability considering the two classes problem. Analogously for CNN2classes2ndClass, and for three classes classification with $CNN3classes1stClass$, $CNN3classes2ndClass$ and $CNN3classes3rdClass$.
 \item $CNN2classes1stClassPerc$ is the probabibility percentage of the 1st class in the two classes problem. Analogously for $CNN2classes2ndClassPerc$,  and for three classes classification with $CNN3classes1stClassPerc$, $CNN3classes2ndClassPerc$ and $CNN3classes3rdClassPerc$.
\end{itemize}

\begin{figure*}
  \centering
  \includegraphics[height=0.96\textheight]{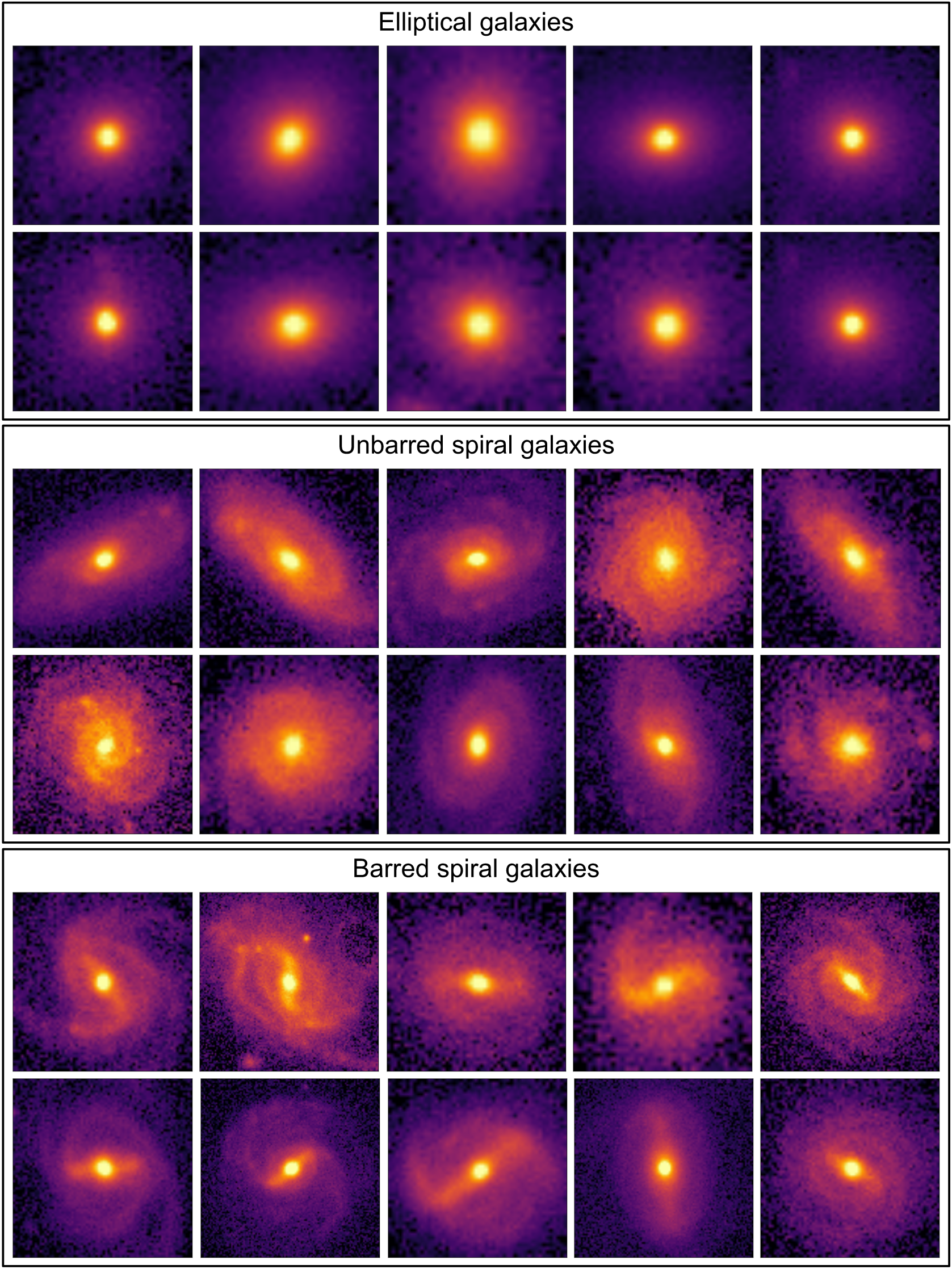}
  \caption{A sample of galaxies classified here in this work using Deep Learning, 
   Galaxy Zoo 2 as supervision (3 classes problem) from $K\ge10$ dataset.  Elliptical galaxies (two top rows), unbarred spirals (two middle rows), and barred spirals (two bottom rows).}
  \label{fig:DL_3classes_k10}
\end{figure*}

Both classifications for two classes problems are performed by models trained with imbalanced datasets. For the 3 classes separation, we use the model trained with the SMOTE dataset. For classifications provided by CNN, we use models trained with $K \ge 5$ dataset. On the one hand, it is important to remember that TML does not classify galaxies with some problem detected by CyMorph. On the other hand, DL acts directly upon images and has no error detection. This catalog represents a significant improvement for extragalactic studies related to galaxy morphologies. The Galaxy Zoo project \citep[GZ, ][]{GZ1a,GZ1b,GZ2} is a great success in offering large numbers of galaxies with reliable morphological classification. Nevertheless, GZ does not provide morphology information for a significant fraction of galaxies. Our catalog complement such effort. We show in Subsection \ref{sub:case_study} that such classification is especially relevant in sparsely inhabited areas of the colour-magnitude diagram, i.e. red spirals and blue ellipticals, for which the lack of objects renders the measurement imprecise. In Figure \ref{fig:DL_3classes_k10} we exhibit some typical examples of each class, where we can see the high quality of the classification for large objects ($K \ge 10$).

\section*{References}

\bibliography{all_references}

\end{document}